\documentclass[showpacs,preprintnumbers,prd,nofootinbib,floats,amssymb,floatfix]{revtex4}
\usepackage{amsmath}
\usepackage{amsfonts}
\usepackage{graphicx}
\usepackage{hyperref}

\usepackage{amsmath}
\usepackage{amstext}
 
\begin{document}
\title{Faddeev-Jackiw quantization of four dimensional BF theory }
\author{ Alberto Escalante} \email{aescalan@ifuap.buap.mx}
 \affiliation{  Instituto de F{\'i}sica, Universidad Aut\'onoma de Puebla,  \\
 Apartado Postal J-48 72570, Puebla Pue., M\'exico. }
\author{Prihel Cavildo S\'anchez} \email{pcavildo@ifuap.buap.mx} 
\affiliation{  Instituto de F{\'i}sica, Universidad Aut\'onoma de Puebla,  \\
 Apartado Postal J-48 72570, Puebla Pue., M\'exico. }

\begin{abstract}
The symplectic analysis of a four dimensional $BF$ theory  in the context of the Faddeev-Jackiw symplectic approach is performed.  It is shown that this method  is more economical than   Dirac's  formalism. In particular, the complete set of  Faddeev-Jackiw constraints  and the generalized Faddeev-Jackiw brackets are reported. In addition, we show that the generalized Faddeev-Jackiw brackets and the Dirac ones coincide to each other. Finally, the similarities and advantages between  Faddeev-Jackiw method and   Dirac's formalism  are briefly discussed.
\end{abstract}
\date{\today}
\preprint{}
\maketitle
\section{INTRODUCTION}
It is well-known that  the topological theories have a relevant role in the context of gravity. In fact, topological theories are  good laboratories for testing classical and quantum ideas of generally covariant gauge systems. Topological theories are characterized by laking of physical  degrees of freedom, either in three or  four dimensions they have a close relation with General Relativity [GR]  just as the  background independence  and the diffeomorphisms covariance, this is, all the dynamical variables characterizing  the theory are dynamical ones.  In the three dimensional case a relevant example  of   topological theory   is  the  Chern-Simons theory. In fact,  basically Chern-Simons theory  describes GR,  it has been showed that these theories are equivalent up to a total derivative \cite{1, 2},  and also there exist a relation between these theories defined with  (or  without) an Immirizi-like  parameter \cite {3, 4}. Furthermore,  we can find a recent work where the Chern-Simons state describes a topological state with unbroken diffeomorphism invariance in Yang-Mills and GR \cite{5}. In the Loop Quantum Gravity context, that state is called the Kodama state and has been studied in interesting works by Smolin, arguing that the Kodama state at least for the de  Sitter spacetime, Loop Quantum Gravity does have a good low energy limit \cite{6}. On the other hand,  in  four dimensions   there exist  the so-called  $BF$ theory. In fact, $BF$ theories were introduced as generalizations of three dimensional Chern-Simons actions or in other cases, can also be considered  as a zero coupling limit of Yang-Mills theories \cite{7, 8}.  Moreover,  we find in the literature several examples where $BF$ theories with additional  extra constraints describe gravity, for instance,   the well-known  formulations of  Plebanski  and  Macdowell-Mansouri  \cite{9,10}. In addition,    within  the modern  quantization  scheme  using   tools developed in Loop Quantum Gravity,  $BF$  theories have  been studied in the context of spin foams. In fact, in this approach  is not considered the   traditional Fock space formalism but    holonomies along paths as the basic variables to be quantized \cite{11}. With respect  the classical context, there are several works studying the canonical structure of a $BF$ theory,  see  the for instance the references \cite{12,13, 14c}. However, in these works has been used the canonical formalism by using a reduced phase space, this means, it  has been considered as dynamical variables only those that occur in the Lagrangian density with temporal derivative, however, in several cases this approach is not convenient, for instance in Palatini's theory   the price to pay for developing the standard approach is that we cannot know the full structure of the constraints and their algebra is not closed \cite{14}, thus, the better way to carry out the canonical formalism is by following all  Dirac's  steps as it has been commented in  \cite{14a, 15, 16, 17, 18, 19}. In consideration with  the commented above,  either the classical or the quantum study of  $BF$ theories and their close relation with GR   is at the present a frontier subject of study   \cite{20, 21}. \\
In this manner, with  the ideas explained previously  in this work the Faddeev-Jackiw [FJ] symplectic quantization of a   four-dimensional $BF$ theory  is performed. In fact,  the FJ method provides an alternative  approach for studying constrained systems based on a first-order Lagrangian \cite{22, 23}.  The FJ method is a symplectic study and the basic feature of this approach is to treat all the constraints at the same footing. In other words,  in   FJ method   one avoids the classification of the constraints into first-class and second-class ones as in Dirac's framework is done. In addition, some essential elements of a physical theory such as the degrees of freedom, the gauge symmetry and the quantization brackets called the  generalized  FJ brackets can also be derived; Dirac's and generalized FJ brackets coincide to each other. However,  it is important to remark that  in  the  canonical  formalism  we must to   work by following all   Dirac's  steps in oder to compare  with the  FJ symplectic formalism. In fact, it  has been showed that by following all Dirac's  steps, the Dirac results and the FJ ones coincide \cite{23a}.  In this respect, in this paper  we also  develop   a  pure canonical analysis and we compare the obtained  results  with the FJ ones. We will start with a $SO(3, 1)$ invariant four-dimensional $BF$ theory, however, we will  break down  the Lorentz group  in order to work with a compact group,  the remaining group will be $SO(3)$. It is important to comment that in \cite{ 24} a pure canonical analysis of a  $SO(3, 1)$  invariant $BF$ theory has been performed, however, in that paper the Dirac brackets were not reported. The reason is that by working with the $SO(3, 1)$ group either the Dirac  or FJ  constraints of the theory have not a simple  structure and this fact difficults the construction of such brackets. In this respect,  we report the  complete structure of the constraints of the theory, then   the Dirac and the  generalized FJ brackets are computed, we will show that the Dirac brackets and the FJ ones coincide to each other.  In this manner,  our results complete and  extend those reported in the literature.\\
The paper is organized as follows. In Section II the FJ analysis for a four-dimensional $BF$ theory is performed; we report the complete set of FJ constraints. Moreover, in order to obtain a symplectic tensor we fix the gauge, then the generalized FJ brackets are found. In Section III, we develop a pure canonical analysis of the theory under study. We report the complete structure  of the first class  and second class constraints  and we show that the algebra between the constraints is closed in full agreement with the canonical rules of Dirac's formulation.  Then  by introducing the Dirac brackets we eliminate the second class constraints. In Section IV we present some remarks and conclusions.

\section{Faddeev-Jackiw Framework  for   BF theory}
In this section we shall perform the FJ analysis, our laboratory will be given by a four-dimensional  $BF$ theory  described by  the following action 
\begin{eqnarray}
S [A_{\mu}^{ IJ}, B_{\alpha\beta}^{ KL}] = \Xi\int_M F^{IJ}\wedge B_{IJ},
\end{eqnarray} 
where $\Xi$ is a constant, $B^{IJ}=\frac{1}{2} B^{IJ}_{\alpha \beta} dx^\alpha \wedge dx^\beta $ is a set of six $SO(3,1)$ valued two forms, the two-form curvature $F$ of the Lorentz connection is defined as usual by  $F_{\mu\nu IJ} = \partial_{\mu}A_{\nu IJ} - \partial_{\nu}A_{\mu IJ} + A_{\mu IK}A^{\;K}_{\nu\;\;J} - A_{\mu JK}A^{\;K}_{\nu\;\;I}$. Here, $I,J,K... = 0,1,2,3$ are internal Lorentz  indices  that can be raised and lowered by the internal metric $\eta_{IJ}=(-1, 1, 1, 1)$,   $x^\mu$ are the coordinates that label the points of  the four-dimensional manifold $M$,  and $\alpha, \beta, \mu, ... = 0,1,2,3$  are space-time indices. \\
By performing the  $3+1$ decomposition and breaking the Lorentz group down to $SO(3)$    we obtain the following Lagrangian density 
\begin{eqnarray}
L &=& \Xi\int \eta^{abc}\left[ B_{ab}^{\;\;\;0i}\dot{A}_{c0i} + \frac{1}{2}B_{ab}^{\;\;\;ij}\dot{A}_{cij} \right.\nonumber \\
&& + \frac{1}{2}A_{0ij}\left( \partial_{c}B_{ab}^{\;\;\;ij} + B_{ab}^{\,\;\;il}A_{c\;\;l}^{\;\;j} + 2B_{ab}^{\,\;\;0i}A_{c0}^{\;\;\;j}\right)\nonumber\\
&& + A_{00i}\left( \partial_{c}B_{ab}^{\,\;\;0i} + B_{ab}^{\;\;\;0j}A_{c\;\;j}^{\;\;i} + B_{ab}^{\;\;\;ij}A_{c0j} \right) \nonumber\\
&& + B_{0a}^{\;\;\;0i}\left( \partial_{b}A_{c0i} - \partial_{c}A_{b0i} + A_{b0j}A_{c\;\;i}^{\;\;j} + A_{c0}^{\;\;\;j}A_{bij} \right) \nonumber\\
&&\left.  + \frac{1}{2}B_{0a}^{\;\;\;ij}\left( \partial_{b}A_{cij} - \partial_{c}A_{bij} + A_{bil}A_{c\;\;j}^{\;\;l} + A_{bi0}A_{c\;\;j}^{\;\;0} - A_{bjl}A_{c\;\;i}^{\;\;l} - A_{bj0}A_{c\;\;i}^{\;\;0} \right)\right] d^{3}x,
\label{2a}
\end{eqnarray}
here,  $a,b,c,...= 1,2,3$, $\epsilon^{0abc}=\eta^{abc}$ and $i,j,k,l...=1,2,3$ are lowered and raised with the Euclidean metric $\eta_{ij}=(1, 1, 1)$. By introducing the following variables \cite{25}
\begin{eqnarray}
A_{aij} &\equiv & -\epsilon_{ijk}A_{a}^{\;\;k},\nonumber \\
A_{0ij} &\equiv & -\epsilon_{ijk}A_{0}^{\;\;k}, \nonumber \\
B_{abij} &\equiv & -\epsilon_{ijk}B_{ab}^{\;\;\;k},\nonumber \\
B_{0aij} &\equiv & -\epsilon_{ijk}B_{0a}^{\;\;\;k}, \nonumber \\
A_{ai} &\equiv & \Upsilon_{ai}, 
\label{eq3a}
\end{eqnarray}
the  Lagrangian  takes  the following form 
\begin{eqnarray}
\mathop{\mathcal{L}}^{} &=& \Xi\eta^{abc}B_{ab}^{\;\;\;0i}\dot{A}_{c0i} + \Xi\eta^{abc}B_{abi}\dot{\Upsilon}_{c}^{\;\;i} \\
&& - \left[ - A_{0}^{\;\;i}\left( \partial_{c}\left( \Xi\eta^{abc}B_{abi}\right) + \Xi\eta^{abc}\epsilon^{j}_{\;\;ik}B_{abj}\Upsilon_{c}^{\;\;k} - \Xi\eta^{abc}\epsilon_{jki}B_{ab}^{\;\;\;0j}A_{c0}^{\;\;\;k}  \right) \right. \nonumber\\
&& -A_{00i}\left( \partial_{c}\left( \Xi\eta^{abc}B_{ab}^{\;\;\;0i} \right) -\Xi\eta^{abc}\epsilon^{i}_{\;\;jk}B_{ab}^{\;\;\;0j}\Upsilon_{c}^{\;\;k} - \Xi\eta^{abc}\epsilon^{ijk}B_{abk}A_{c0j}  \right)\nonumber\\
&& - \Xi\eta^{abc}B_{0a}^{\;\;\;0i}\left( \partial_{b}A_{c0i} - \partial_{c}A_{b0i} + \epsilon_{i}^{\;\;jk}A_{b0j}\Upsilon_{ck} - \epsilon_{ijk}A_{c0}^{\;\;\;j}\Upsilon_{b}^{\;\;k} \right) \nonumber\\
&&\left.  - \Xi\eta^{abc}B_{0ai}\left( \partial_{b}\Upsilon_{c}^{\;\;i} - \partial_{c}\Upsilon_{b}^{\;\;i} + \epsilon^{i}_{\;\;jk}\Upsilon_{b}^{\;\;j}\Upsilon_{c}^{\;\;k} - \epsilon^{ijk}A_{b0j}A_{c0k} \right)\right] .
\end{eqnarray}
In this manner, we can identify the symplectic Lagrangian given by 
\begin{eqnarray}
\mathop{\mathcal{L}}^{(0)} &=& \Xi\eta^{abc}B_{ab}^{\;\;\;0i}\dot{A}_{c0i} + \Xi\eta^{abc}B_{abi}\dot{\Upsilon}_{c}^{\;\;i} - \mathop{\mathcal{V}}^{(0)},
\label{eq6a}
\end{eqnarray}
where $\mathop{\mathcal{V}}\limits^{(0)}$ is the symplectic potential expressed as 
\begin{eqnarray}
\mathop{\mathcal{V}}^{(0)}&=&- A_{0}^{\;\;i}\left( \partial_{c}\left( \Xi\eta^{abc}B_{abi}\right) + \Xi\eta^{abc}\epsilon^{j}_{\;\;ik}B_{abj}\Upsilon_{c}^{\;\;k} - \Xi\eta^{abc}\epsilon_{jki}B_{ab}^{\;\;\;0j}A_{c0}^{\;\;\;k}  \right) \nonumber\\
&& -A_{00i}\left( \partial_{c}\left( \Xi\eta^{abc}B_{ab}^{\;\;\;0i} \right) -\Xi\eta^{abc}\epsilon^{i}_{\;\;jk}B_{ab}^{\;\;\;0j}\Upsilon_{c}^{\;\;k} - \Xi\eta^{abc}\epsilon^{ijk}B_{abk}A_{c0j}  \right)\nonumber\\
&& - \Xi\eta^{abc}B_{0a}^{\;\;\;0i}\left( \partial_{b}A_{c0i} - \partial_{c}A_{b0i} + \epsilon_{i}^{\;\;jk}A_{b0j}\Upsilon_{ck} - \epsilon_{ijk}A_{c0}^{\;\;\;j}\Upsilon_{b}^{\;\;k} \right) \nonumber\\
&& - \Xi\eta^{abc}B_{0ai}\left( \partial_{b}\Upsilon_{c}^{\;\;i} - \partial_{c}\Upsilon_{b}^{\;\;i} + \epsilon^{i}_{\;\;jk}\Upsilon_{b}^{\;\;j}\Upsilon_{c}^{\;\;k} - \epsilon^{ijk}A_{b0j}A_{c0k} \right).
\end{eqnarray}
From the symplectic Lagrangian  (\ref{eq6a}) we identify the following symplectic variables $\mathop{\mathcal{  \xi}}^{(0)} = \left( A_{a0i}, B_{ab}^{\;\;\;0i}, \Upsilon_{a}^{\;\;i}, B_{abi}, A_{0}^{\;\;i}, A_{00i},B_{0a}^{\;\;\;0i}, B_{0ai} \right)$
and the 1-forms $ \mathrm{a}^{(0)} = \left( \Xi\eta^{abc}B_{ab}^{\;\;\;0i},0,\Xi\eta^{abc}B_{abi},0,0,0,0,0 \right)$.
In this manner, the symplectic matrix defined as $ f^{(0)}_{ij}(x,y)=\frac{\delta \mathrm{a}_{j}(y)}{\delta\xi^{i}(x)}-\frac{\delta \mathrm{a}_{j}(x)}{\delta\xi^{i}(y)},
$ is given by
\begin{eqnarray}
\mathop{f}^{(0)}\ _{ij} &=& \left(
\begin{array}{cccccccc}
0  & -\Xi\eta^{abc}\delta_{j}^{i} & 0 & 0 & 0 & 0 & 0 & 0 \\
\Xi\eta^{abc}\delta_{j}^{i} & 0 & 0 & 0 & 0 & 0 & 0 & 0  \\
0 & 0 & 0 & -\Xi\eta^{abc}\delta_{j}^{i} & 0 & 0 & 0 & 0 \\
0 & 0 & \Xi\eta^{abc}\delta^{i}_{j} & 0 & 0 & 0 & 0 & 0 \\
0 & 0 & 0 & 0 & 0 & 0 & 0 & 0 \\
0 & 0 & 0 & 0 & 0 & 0 & 0 & 0 \\
0 & 0 & 0 & 0 & 0 & 0 & 0 & 0 \\
0 & 0 & 0 & 0 & 0 & 0 & 0 & 0 
\end{array}
\right)\delta^{3}(x-y),
\end{eqnarray}
we observe that $\mathop{f}^{(0)}_{ij}$ is singular and therefore, there will constraints. The modes of $\mathop{f}\limits^{(0)}\ _{ij}$ are given by the following 4 vectors 
\begin{eqnarray}
v^{(0)}\ _{1}^{} &=& \left( 0,0,0,0,V^{A_{0}^{i}},0,0,0 \right),\\ 
v^{(0)}\ _{2}^{} &=& \left( 0,0,0,0,0,V^{A_{00i}},0,0 \right),\\
v^{(0)}\ _{3}^{} &=& \left( 0,0,0,0, 0,0, V^{B_{0a}^{0i}},0 \right),\\  
v^{(0)}\ _{4}^{} &=& \left( 0,0,0,0,0, 0,0, V^{B_{0ai}}\right),
\end{eqnarray}
where $V^{A_{0}^{i}}, V^{A_{00i}},  V^{B_{0a}^{0i}}$  and $V^{B_{0ai}}$are arbitrary functions.  Hence, by using these modes we find the following constraints
\begin{eqnarray}
\mathop{\Omega}\limits^{(0)}\  _{i} &=& \int d^{3}x v^{(0)}\ _{1}^{i}\frac{\delta}{\delta \xi^{i}}\int d^{3}y \mathop{\mathcal{V}}^{(0)}(\xi)\nonumber \\
&=&\int d^{3}x V^{A_{0}^{i}}\frac{\delta}{\delta A_{0}^{\;\;i}}\int d^{3}y \mathop{\mathcal{V}}^{(0)}(\xi)\nonumber\\
&=& \partial_{c}\left( \Xi\eta^{abc}B_{abi} \right) + \Xi\eta^{abc}\epsilon^{j}_{\;\;ik}B_{abj}\Upsilon^{k}_{c} - \Xi\eta^{abc}\epsilon_{jki}B_{ab}^{\;\;\;0j}A_{c0}^{\;\;k},\nonumber \\
\mathop{\Omega}^{(0)}\ ^{00i} &=& \int d^{3}x v^{(0)}\ _{2}^{i}\frac{\delta}{\delta \xi^{i}}\int d^{3}y \mathop{\mathcal{V}}^{(0)}(\xi)\nonumber\\
&=&\int d^{3}x V^{A_{00i}}\frac{\delta}{\delta A_{00i}}\int d^{3}y \mathop{\mathcal{V}}^{(0)}(\xi)\nonumber\\
&=& \partial_{c}\left( \Xi\eta^{abc}B_{ab}^{\;\;\;0i} \right) - \Xi\eta^{abc}\epsilon^{i}_{\;\;jk}B_{ab}^{\;\;\;0j}\Upsilon^{k}_{c} - \Xi\eta^{abc}\epsilon^{ijk}B_{abk}A_{c0j},\nonumber \\
\mathop{\Omega}^{(0)}\ ^{0a}_{\;\;\;0i} &=& \int d^{3}x v^{(0)}\ _{3}^{i}\frac{\delta}{\delta \xi^{i}}\int d^{3}y \mathop{\mathcal{V}}^{(0)}(\xi)\nonumber\\
&=&\int d^{3}x V^{B_{0a}^{\;\;\;0i}}\frac{\delta}{\delta B_{0a}^{\;\;\;0i}}\int d^{3}y \mathop{\mathcal{V}}^{(0)}(\xi)\nonumber\\
&=& \Xi \eta^{abc}\left( \partial_{b}A_{c0i} - \partial_{c}A_{b0i} + \epsilon_{i}^{\;\;jk}A_{b0j}\Upsilon_{ck} - \epsilon_{ijk}A_{c0}^{\;\;\;j}\Upsilon_{b}^{\;\;k}  \right) ,\nonumber \\
\mathop{\Omega}^{(0)}\ ^{0ai} &=& \int d^{3}x v^{(0)}\ _{4}^{i}\frac{\delta}{\delta \xi^{i}}\int d^{3}y \mathop{\mathcal{V}}^{(0)}(\xi)\nonumber\\
&=&\int d^{3}x V^{B_{0ai}}\frac{\delta}{\delta B_{0ai}}\int d^{3}y \mathop{\mathcal{V}}^{(0)}(\xi)\nonumber\\
&=& \Xi\eta^{abc}\left( \partial_{b}\Upsilon_{c}^{\;\;i} - \partial_{c}\Upsilon_{b}^{\;\;i} + \epsilon^{i}_{\;\;jk}\Upsilon_{b}^{\;\;j}\Upsilon_{c}^{\;\;k} - \epsilon^{ijk}A_{b0j}A_{c0k} \right), 
\label{cons}
\end{eqnarray}
we can observe that these constraints are the  secondary constraints obtained by using the Dirac method (see the  following section).
Now we shall observe if there are more constraints, for this aim, we calculate the following system \cite{26}
\begin{eqnarray}
\bar{f}_{kj}^{}\dot{\xi}^{(0)j}=Z_{k}^{}(\xi),
\label{p}
\end{eqnarray}
where
\begin{eqnarray}
\bar{f}_{kj}^{}=\left(
\begin{array}{cc}
f^{(0)}_{ij} \\
\frac{\delta\Omega^{(0)}_i}{\delta\xi^{(0)j}}
\end{array}\right)\hphantom{111}\mathrm{and}\hphantom{111}Z_{k}^{}=
\left(
\begin{array}{cc}
\frac{\delta \mathop{\mathcal{V}}^{(0)}}{\delta\xi^{(0)j}} \\
0\\
0\\
0
\end{array}
\right).
\end{eqnarray}
Thus,  the symplectic matrix $\bar{f}_{ij}^{}$ is given by
\begin{eqnarray}
\bar{f}_{ij} &=&\left( \begin{array}{cccccccc}
0 & -\Xi \eta^{abc}\delta_{j}^{i} & 0  \\
\Xi\eta^{abc}\delta_{j}^{i} & 0 & 0  \\
0 & 0 & 0  \\
0 & 0 & \Xi\eta^{abc}\delta_{j}^{i}  \\
0 & 0 & 0  \\
0 & 0 & 0  \\
0 & 0 & 0   \\
0 & 0 & 0  \\
-\Xi\eta^{abc}\epsilon_{jki}B_{ab}^{\;\;\;oj} & -\Xi\eta^{abc}\epsilon_{kji}A_{c0}^{\;\;\;j} & \Xi\eta^{abc}\epsilon^{j}_{\;\;ik}B_{abj} \\
-\Xi\eta^{abc}\epsilon^{ikj}B_{abj} & \Xi\eta^{abc}\left( \delta_{k}^{i}\partial_{c} - \epsilon^{i}_{\;\;jk}\Upsilon_{c}^{j}  \right)  & -\Xi\eta^{abc}\epsilon^{i}_{\;\;jk}B_{ab}^{\;\;\;0j} \\
2\Xi\eta^{abc}\left( \delta_{i}^{k}\partial_{c} - \epsilon_{i}^{\;\;kj}\Upsilon_{bj} \right)  & 0 & 2\Xi\eta^{abc}\epsilon_{i}^{\;\,jk}A_{b0j} \\
-2\Xi\eta^{abc}\epsilon^{ijk}A_{b0j} & 0 & 2\Xi\eta^{abc}\left(\delta_{k}^{i}\partial_{b} + \epsilon^{i}_{\;\;jk}\Upsilon_{b}^{j}  \right) \\
\end{array}\right.\nonumber\\
 &&\left. \begin{array}{cccccccc}
 0 & 0 & 0 & 0 & 0 \\
 0 & 0 & 0 & 0 & 0 \\
 -\Xi\eta^{abc}\delta_{j}^{i} & 0 & 0 & 0 & 0 \\
 0 & 0 & 0 & 0 & 0 \\
 0 & 0 & 0 & 0 & 0 \\
 0 & 0 & 0 & 0 & 0 \\
 0 & 0 & 0 & 0 & 0 \\
 0 & 0 & 0 & 0 & 0 \\
 \Xi\eta^{abc}\left( \delta_{i}^{k}\partial_{c} + \epsilon^{k}_{\;\,ij}\Upsilon_{c}^{j} \right)  & 0 & 0 & 0 & 0 \\
 -\Xi\eta^{abc}\epsilon^{ijk}A_{c0j} & 0 & 0 & 0 & 0 \\
 0 & 0 & 0 & 0 & 0 \\
 0 & 0 & 0 & 0 & 0 \\
\end{array}\right) \delta^3(x-y),
\end{eqnarray}

The matrix $f_{ij}$ is not a square matrix as expected, however it has null vectors. The null vectors are given by
\begin{eqnarray}
\vec{V}_{1} &=& \left( \epsilon_{kji}A_{c0}^{\;\;\;j}V^{i}, -\epsilon_{jki}B_{ab}^{\;\;\;0j}V^{i}, \partial_{c}V^{k} + \epsilon^{k}_{\;\;ij}\Upsilon_{c}^{i}V^{j}, \epsilon^{j}_{\;\,ik}B_{abj}V^{i}, 0,0 ,0 ,0 , V^{i}, 0, 0, 0   \right), \\
\vec{V}_{2} &=& \left( \partial_{c}V_{k} - \epsilon^{i}_{\;\;kj}\Upsilon_{c}^{j}V_{i}, \epsilon^{ikj}B_{abj}V_{i}, \epsilon^{kij}A_{c0j}V_{i}, \epsilon^{k}_{\;\,ji}B_{ab}^{\;\;\;0j}V^{i}, 0,0,0,0,0, V_{i}, 0, 0  \right), \\ 
\vec{V}_{3} &=& \left( 0, 2\left( \partial_{b}V^{k} - \epsilon_{i}^{\;\;kj}\Upsilon_{bj}V^{i} \right), 0, 2\epsilon_{i}^{\;\;jk}A_{b0j}V^{i}, 0,0,0,0,0,0, V^{i}, 0  \right),\\
\vec{V}_{4} &=& \left( 0, 2\epsilon^{ijk}A_{b0j}V_{i}, 0, 2\left( \partial_{b}V^{k} + \epsilon^{k}_{\;\,ji}\Upsilon_{b}^{j}V^{i}\right), 0,0,0,0,0,0,0, V_{i}  \right). 
\end{eqnarray}
On the other hand, $Z_{k}(\xi) $ is given by 
\begin{eqnarray}
Z_{k}(\xi) &=& \left( \begin{array}{c}
\frac{\delta \mathop{\mathcal{V}}\limits^{(0)}(\xi)}{\delta \xi^{i}}\nonumber\\
0
\end{array}\right) = \left( \begin{array}{c}
\Xi \eta^{abc}\epsilon_{j\;\;i}^{\;\;k}A_{0}^{\;\,i}B_{ab}^{\;\,\;0j} + \eta^{abc}\epsilon^{ikj}A_{00i}B_{abj} + 2\Xi\eta^{abc}\partial_{b}B_{0a}^{\;\;\,0k}  \nonumber\\
- 2\Xi\eta^{abc}\epsilon_{i}^{\;\,kj}B_{0a}^{\,\,0i}\Upsilon_{cj}  + 2\Xi\eta^{abc}\epsilon^{ijk}B_{0ai}A_{b0j}\nonumber\\\\
\Xi\eta^{abc}\epsilon_{ikj}A_{0}^{\;\;j}A_{c0}^{\;\;\,k} + \Xi\eta^{abc}\partial_{c}A_{00i} + \Xi\eta^{abc}\epsilon^{j}_{\;\;ik} A_{00j}\Upsilon_{c}^{k}\nonumber\\\\
-\Xi\eta^{abc}\epsilon^{j}_{\;\;ik}A_{0}^{\;\;i}B_{abj} + \Xi\eta^{abc}\epsilon^{i}_{\;\;jk}A_{00i}B_{ab}^{\;\;\;0j} - 2\Xi\eta^{abc}\epsilon_{i}^{\;\,jk}A_{b0j}B_{0a}^{\;\;0i}  \nonumber\\
- 2\Xi\eta^{abc}\partial_{b}B_{0ak} - 2\epsilon^{i}_{\;\;jk}\Upsilon_{b}^{j}B_{0ai}\\\\
\Xi\eta^{abc}\partial_{c}A_{0}^{\;\;\;i} - \Xi\eta^{abc}\epsilon^{i}_{\;\;jk}A_{0}^{\;\,\;j}\Upsilon_{c}^{k}
 + \Xi\eta^{abc}\epsilon^{jki}A_{00j}A_{c0k}\\\\
 \mathop{\Omega}\limits^{(0)}\ _{i}\\\\
 \mathop{\Omega}\limits^{(0)}\ ^{00i}\\\\
 \mathop{\Omega}\limits^{(0)}\ ^{0a}_{\;\;\;0i}\\\\
 \mathop{\Omega}\limits^{(0)}\ ^{0ai}\\\\
 0\\\\
 0\\\\
 0\\\\
 0
 \end{array} \right)  
\end{eqnarray}

The contraction of the null vectors with $Z_{k}$, namely, $\vec{V}_{i}^{\mu}Z_{\mu}(\xi) = 0$, give identities. For instance, from  the contraction of $\vec{V}_{1}$ with $Z_{k}(\xi)$ we obtain
\begin{eqnarray}
\vec{V}_{1}^{\mu}Z_{\mu}(\xi) &=& \epsilon^{j}_{\;\,ik}A_{0}^{\;\;i}V^{k}\left[ \partial_{c}\left( \Xi\eta^{abc}B_{abj}\right) + \Xi\eta\epsilon^{m}_{\;\,jkl}B_{abm}\Upsilon_{c}^{l} + \Xi\eta^{abc}\epsilon_{plj}B_{ab}^{\;\,\,0p}A_{c0}^{\,\,\,l}  \right] \nonumber\\
&& + \epsilon^{i}_{\;\,kj}A_{00i}V^{k}\left[ \Xi\eta^{abc}\partial_{c}B_{ab}^{\;\,\,0j} - \Xi\eta^{abc}\epsilon^{j\;\,m}_{\;\,i}B_{abm}A_{c0}^{\;\;\;l} - \Xi\eta\epsilon^{j}_{\;\,nl}B_{ab}^{\;\,\,0n}\Upsilon_{c}^{l} \right]\nonumber\\ 
&& - \epsilon^{i}_{\;\,jk}B_{0ai}V^{k}\left[ \Xi\eta^{abc}\left( \partial_{b}\Upsilon_{c}^{j} - \partial_{c}\Upsilon_{b}^{j} \right) + \Xi\eta^{abc}\epsilon^{j}_{\;\,pl}\Upsilon_{b}^{p}\Upsilon_{c}^{l} - \Xi\eta^{abc}\epsilon^{j}_{\;\,pl}A_{b0}^{\;\;\;p}A_{c0}^{\;\,\,l}  \right] \nonumber\\
&& - \epsilon_{ijk}B_{0a}^{\;\;\,0i}V^{k}\left[ \Xi\eta^{abc}\left(\partial_{b}A_{c0}^{\;\;\;j} - \partial_{c}A_{b0}^{\;\,\,j} \right) + \epsilon^{j}_{\;\,ml}\Upsilon_{c}^{l}A_{b0}^{\;\;\;m} - \epsilon_{jml}\Upsilon_{b}^{l}A_{c0}^{\;\,\;m}  \right]\nonumber\\
&=& 0,
\end{eqnarray}
where we  can observe that the left hand side vanishes  because  is a linear combination of constraints. Hence,  there are not more FJ  constraints.\\
Furtheremore, we will add the constraints given in (\ref{cons}) to  the symplectic Lagrangian  using the following Lagrange multipliers, namely, $A_{0}^{\;\;i} = \dot{T}^{i}, A_{00i} = \dot{\Lambda}_{i}, B_{0a}^{\;\;\;0i} = \frac{\dot{\varsigma}_{a}^{i}}{2}, B_{0ai} = \frac{\dot{\chi}_{ai}}{2}$, thus the symplectic Lagrangian takes the form
\begin{eqnarray}
\mathop{\mathcal{L}}^{(1)} &=& \Xi\eta^{abc}B_{ab}^{\;\;\;0i}\dot{A}_{c0i} + \Xi\eta^{abc}B_{abi}\dot{\Upsilon}_{c}^{\;\;i} - \dot{T}^{i}\mathop{\Omega}^{(0)}\ _{i} - \dot{\Lambda}_{i}\mathop{\Omega}^{(0)}\ ^{00i} - \frac{\dot{\varsigma}_{a}^{\;\;i}}{2}\mathop{\Omega}^{(0)}\ ^{0a}_{\;\;\;0i}\nonumber\\
&& - \frac{\dot{\chi}_{ai}}{2}\mathop{\Omega}^{(0)}\ ^{0ai} - \mathop{\mathcal{V}}^{(1)},
\label{23a}
\end{eqnarray}
where $\mathop{\mathcal{V}}^{(1)} = \mathop{\mathcal{V}}^{(0)}\mid_{\mathop{\Omega}\limits^{(0)}\ _{i}, \mathop{\Omega}\limits^{(0)}\ ^{00i}, \mathop{\Omega}\limits^{(0)}\ ^{0a}_{\;\;\;0i}, \mathop{\Omega}\limits^{(0)}\ ^{0ai} =0} =0,$ this result is expected because of the general covariance of the theory just as it  is present in General Relativity. \\
From the symplectic Lagrangian (\ref{23a}) we identify the following symplectic variables $\mathop{\xi}^{(1)} = \left( A_{c0i}, B_{ab}^{\;\,\,0i}, \Upsilon_{c}^{i}, B_{abi}, T^{i}, \Lambda_{i}, \varsigma_{a}^{\,\;i}, \chi_{ai} \right) $
and the 1 - forms $\mathrm{a}^{(1)} = \left( \Xi\eta^{abc}B_{ab}^{\;\,\,0i}, 0, \Xi\eta^{abc}B_{abi}, 0, - \mathop{\Omega}^{(0)}\ _{i}, -\mathop{\Omega}^{(0)}\ ^{00i}, -\frac{\mathop{\Omega}\limits^{(0)}\ ^{0a}_{\;\;\;0i}}{2}, -\frac{\mathop{\Omega}\limits^{(0)}\ ^{0ai}}{2} \right)$. Hence, the symplectic matrix  has the following form 
\begin{eqnarray}
\mathop{f}^{(1)}\ _{ij} &=& \left(\begin{array}{cccccccc}
0 & -\Xi\eta^{abc}\delta_{j}^{i} & 0 & 0 \\
\Xi\eta^{abc}\delta_{j}^{i} & 0 & 0 & 0 \\	
0 & 0 & 0 & -\Xi\eta^{abc}\delta_{j}^{i}  \\
0 & 0 & \Xi\eta^{abc}\delta^{i}_{j} & 0  \\
-\Xi\eta^{abc}\epsilon_{j\;\,i}^{\;\;k}B_{ab}^{\;\;\;0j} & -\Xi\eta^{abc}\epsilon_{kji}A_{c0}^{\;\,\;j} & \Xi\eta^{abc}\epsilon^{j}_{\;\;ik}B_{abj} & \Xi\eta^{abc}D_{c\;\;i}^{\;\;k}  \\
-\Xi\eta^{abc}\epsilon^{ikj}B_{abj} & \Xi\eta^{abc}d_{c\;\;k}^{\;\;i}  & \Xi\eta^{abc}\epsilon^{i}_{\;\;jk}B_{ab}^{\;\;\;0i} & \Xi\eta^{abc}\epsilon^{ijk}A_{c0j} \\
\Xi\eta^{abc}d_{bi}^{\;\;\;k}  & 0 & \Xi\eta^{abc}\epsilon_{i\;\;k}^{\;\;j}A_{b0j} & 0  \\
-\Xi\eta^{abc}\epsilon^{ijk} A_{b0j} & 0 & \Xi\eta^{abc}d_{b\;\;k}^{\;\;i}  & 0  \\
\end{array}  \right. \nonumber\\\nonumber\\
&& \left.\begin{array}{cccccccc}
 \Xi\eta^{abc}\epsilon_{j\;\;i}^{\;\,k}B_{ab}^{\;\;\;0j} & \Xi\eta^{abc}\epsilon^{ikj}B_{abj} & \Xi\eta^{abc}d_{ai}^{\;\;\;k}  & \Xi\eta^{abc}\epsilon^{ijk}A_{b0j} \\
 \Xi\eta^{abc}\epsilon_{kji}A_{c0}^{\;\,\;j} & -\Xi\eta^{abc}d_{c\;\;k}^{\;\;i}  & 0 & 0 \\	
 -\Xi\eta^{abc}\epsilon^{j}_{\;\,ik}B_{abj} & \Xi\eta^{abc}\epsilon^{i}_{\;\,jk}B_{ab}^{\;\,\;0i} & -\Xi\eta^{abc}\epsilon_{i\;\;k}^{\;\;j}A_{b0j} & -\Xi\eta^{abc}D_{b\;\;j}^{\;\;i}  \\
 -\Xi\eta^{abc}D_{c\;\;i}^{\;\;k}  & \Xi\eta^{abc}\epsilon^{ijk}A_{c0j} & 0 & 0 \\
 0 & 0 & 0 & 0 \\
 0 & 0 & 0 & 0 \\
0 & 0 & 0 & 0 \\
 0 & 0 & 0 & 0 \\
\end{array}  \right) \delta^{3}(x-y),  
\end{eqnarray}
 where  we have used the notation $D_{al}^{\;\;\;i} = \delta_{l}^{i}\partial_{a} + \epsilon_{l}^{\;\;ik}\Upsilon_{ak}$ and $d_{ai}^{\;\,\;l} = \delta_{i}^{l}\partial_{a} - \epsilon_{i}^{\;\;lk}\Upsilon_{ak}$.
We can observe that this matrix is singular, however we have showed that there are not more constraints, therefore the theory under study has a gauge symmetry. In order to obtain a symplectic tensor, we fix the following temporal gauge
\begin{eqnarray}
A_{0}^{\;\;i} &=&0,\\
A_{00i} &=& 0, \\
B_{0a}^{\;\;\;0i} &=& 0,\\
B_{0ai} &=& 0,
\end{eqnarray}
this mean that $\dot{T}^{i} =0, \dot{\Lambda}_{i}=0, \dot{\varsigma}_{a}^{i}=0$ and $\dot{\chi}_{ai}=0$. In this manner, we introduce more Lagrange multipliers  enforcing the gauge fixing. The Lagrange multipliers introduced  are $\beta_i, \alpha^i, \rho^a_i, \sigma_a^i$, thus , the symplectic Lagrangian takes the form 
\begin{eqnarray}
\mathop{\mathcal{L}}^{(2)} &=& \Xi\eta^{abc}B_{ab}^{\;\;\;0i}\dot{A}_{c0i} + \Xi\eta^{abc}B_{abi}\dot{\Upsilon}_{c}^{\;\;i} - \left[ \mathop{\Omega}^{(0)}\ _{i} - \beta_{i} \right]\dot{T}^{i} - \left[ \mathop{\Omega}^{(0)}\ ^{00i} - \alpha^{i} \right]\dot{\Lambda}_{i}\nonumber\\
&& - \left[ \frac{\mathop{\Omega}\limits^{(0)}\ ^{0a}_{\;\;\;0i}}{2} - \rho^{a}_{i} \right]\dot{\varsigma}_{a}^{i}   - \left[ \frac{\mathop{\Omega}\limits^{(0)}\ ^{0ai}}{2} - \sigma^{ai} \right]\dot{\chi}_{ai},
\end{eqnarray}
from this symplectic Lagrangian  we identify the following symplectic variables $\mathop{\xi}^{(2)} = \left( A_{a0i}, B_{ab}^{\;\;\;0i}, \Upsilon^{i}_{a}, B_{abi}, T^{i}, \Lambda_{i}, \varsigma_{a}^{i}, \chi_{ai}, \beta_{i}, \alpha^{i}, \rho^{a}_{i}, \sigma^{ai} \right)$,
and the 1-forms
\begin{eqnarray}
\mathrm{a}^{(2)} &=& \left( \Xi\eta^{abc}B_{ab}^{\;\;\;0i},0,\Xi\eta^{abc}B_{abi},0, - \left[ \mathop{\Omega}^{(0)}\ _{i} - \beta_{i} \right], - \left[ \mathop{\Omega}^{(0)}\ ^{00i} - \alpha_{i} \right],\right.\nonumber\\
&& \left. - \left[ \frac{\mathop{\Omega}\limits^{(0)}\ ^{0a}_{0i}}{2} - \rho_{i} \right], - \left[ \frac{\mathop{\Omega}\limits^{(0)}\ ^{0ai}}{2} - \sigma_{i} \right], 0,0,0,0    \right).\nonumber 
\end{eqnarray}
Thus, the symplectic matrix is given by

\begin{eqnarray}
\mathop{f}^{(2)}\ _{ij} &=& \left(
\begin{array}{cccccccccccc}
0  & -\Xi\eta^{abc}\delta^{i}_{j} & 0 & 0 & \Xi\eta^{abc}\epsilon_{j\;\;k}^{\;\;i}B_{ab}^{\;\;\;oj} \\
\Xi\eta^{abc}\delta_{j}^{i} & 0 & 0 & 0 & \Xi\eta^{abc}\epsilon_{ijk}A_{c0}^{\;\;\;j}\\
0 & 0 & 0 & -\Xi\eta^{abc}\delta^{i}_{j} & -\Xi\eta^{abc}\epsilon^{j}_{\;\;\;ki}B_{abj} \\
0 & 0 & \Xi\eta^{abc}\delta_{j}^{i} & 0 & -\Xi\eta^{abc}\epsilon^{i}_{\;\;jk}\Upsilon^{k}_{c} \\
-\Xi\eta^{abc}\epsilon_{j\;\;k}^{\;\;i}B_{ab}^{\;\;\;0j} & -\Xi\eta^{abc}\epsilon_{ijk}A_{c0}^{\;\;\;j} & \Xi\eta^{abc}\epsilon^{j}_{\;\;ki}B_{abj} & \Xi\eta^{abc}\epsilon^{i}_{\;\;jk}\Upsilon^{k}_{c} & 0 \\
-\Xi\eta^{abc}\epsilon^{kij}B_{abj} & \Xi\eta^{abc}d_{c\;\;l}^{\;\;i}   & -\Xi\eta^{abc}\epsilon^{k}_{\;\;ji}B_{ab}^{\;\;\;0j} & -\Xi\eta^{abc}\epsilon^{kji}A_{c0j} & 0 \\
-\Xi\eta^{abc}d_{bl}^{\;\;\;i}   & 0 & -\Xi\eta^{abc}\epsilon_{i}^{\;\;jk}A_{b0j} & 0 & 0 \\
-\Xi\eta^{abc}\epsilon^{kji}A_{b0j} & 0 & \Xi\eta^{abc}d_{b\;\;i}^{\;\;l}  & 0 & 0 \\
0 & 0 & 0 & 0 & \delta_{j}^{i}\\
0 & 0 & 0 & 0 & 0 \\
0 & 0 & 0 & 0 & 0 \\
0 & 0 & 0 & 0 & 0 
\end{array}
\right.\nonumber\\
&& \left.
\begin{array}{cccccccccccc}
\Xi\eta^{abc}\epsilon^{kij}B_{abj} & \Xi\eta^{abc}d_{bj}^{\;\;\;i}  & \Xi\eta^{abc}\epsilon^{kji}A_{b0j} & 0 & 0 & 0 & 0 \\
-\Xi\eta^{abc}d_{c\;\;l}^{\;\;i}   & 0 & 0 & 0 & 0 & 0 & 0  \\
\Xi\epsilon^{k}_{\;\;ji}B_{ab}^{\;\;\;0j} & \Xi\eta^{abc}\epsilon_{i}^{\;\;jk}A_{b0j} & -\Xi\eta^{abc}d_{b\;\;i}^{\;\;l}   & 0 & 0 & 0 & 0 \\
\Xi\eta^{abc}\epsilon^{kji}A_{c0j} & 0 & 0 & 0 & 0 & 0 & 0 \\
 0 & 0 & 0 & -\delta_{j}^{i} & 0 & 0 & 0\\
 0 & 0 & 0 & 0 & -\delta_{j}^{i} & 0 & 0\\
 0 & 0 & 0 & 0 & 0 & -\delta_{b}^{a}\delta_{j}^{i} & 0\\
 0 & 0 & 0 & 0 & 0 & 0 & -\delta_{b}^{a}\delta_{j}^{i}\\
0 & 0 & 0 & 0 & 0 & 0 & 0\\
\delta_{j}^{i} & 0 & 0 & 0 & 0 & 0 & 0\\
 0 & \delta_{b}^{a}\delta_{j}^{i} & 0 & 0 & 0 & 0 & 0\\
 0 & 0 & \delta_{b}^{a}\delta_{j}^{i} & 0 & 0 & 0 & 0
\end{array}
\right)\delta^{3}(x-y),
\end{eqnarray}
where $d_{ai}^{\;\;\;l}\equiv \delta_{i}^{l}\partial_{a} - \epsilon_{i}^{\;\;lk}\Upsilon_{ak}$. We can observe that this matrix is not singular, after a long calculation, the inverse of $\mathop{f}\limits^{(2)}\ _{ij}$ is given by

\begin{eqnarray}
\mathop{f}^{(2)}\ _{ij}^{-1} &=& \left(
\begin{array}{cccccccccccc}
0 & \frac{1}{2\Xi}\eta_{bgc}\delta_{j}^{i} & 0 & 0 & 0 & 0 & 0 & 0 \\
-\frac{1}{2\Xi}\eta_{bgc}\delta_{j}^{i} & 0 & 0 & 0 & 0 & 0 & 0 & 0 \\
0 & 0 & 0 & \frac{1}{2 \Xi}\eta_{abc}\delta_{j}^{i} & 0 & 0 & 0 & 0 \\
0 & 0 & -\frac{1}{2\Xi} \eta_{abc}\delta_{j}^{i} & 0 & 0 & 0 & 0 & 0 \\
0 & 0 & 0 & 0 & 0 & 0 & 0 & 0 \\
0 & 0 & 0 & 0 & 0 & 0 & 0 & 0 \\
0 & 0 & 0 & 0 & 0 & 0 & 0 & 0 \\
0 & 0 & 0 & 0 & 0 & 0 & 0 & 0 \\
\epsilon_{ijl}A_{c0}^{\;\;\;j} & -\epsilon_{j\;\;\;k}^{\;\;l}B_{ab}^{\;\;\;0j} & -\epsilon^{l}_{\;\;jk}\Upsilon_{c}^{k} & -\epsilon^{j}_{\;\;kl}B_{mnj} & -\delta_{j}^{i} & 0 & 0 & 0 \\
0 & -\epsilon^{klj}B_{abj} & \epsilon^{kjl}A_{c0j} & -\frac{1}{2}\delta_{mn}^{ab}\epsilon_{k\;\;l}^{\;\;j}A_{b0j} & 0 & -\delta_{j}^{i} & 0 & 0 \\
0 & -\frac{1}{2}\delta_{mn}^{ab}D_{bi}^{\;\;\;l} & 0 & -\frac{1}{2}\delta_{mn}^{ab}D_{b\;\;l}^{\;\;k} & 0 & 0 & -\delta_{b}^{a}\delta_{j}^{i} & 0 \\
0 & -\frac{1}{2}\delta_{mn}^{ab}\epsilon^{kjl}A_{b0j} & 0 & 0 & 0 & 0 & 0 & -\delta_{b}^{a}\delta_{j}^{i} 
\end{array}
\right.\nonumber\\\nonumber\\
&& \left. 
\begin{array}{cccccccccccc}
-\epsilon_{ijl}A_{c0}^{\;\;\;j} & 0 & 0 & 0\\
 \epsilon_{j\;\;\;k}^{\;\;l}B_{ab}^{\;\;\;0j} & \epsilon^{klj}B_{abj} & \frac{1}{2}\delta_{mn}^{ab}D_{ai}^{\;\;\;l} & \frac{1}{2}\delta_{mn}^{ab}\epsilon^{kjl}A_{boj}\\
 \epsilon^{l}_{\;\;jk}\Upsilon^{k}_{c} & -\epsilon^{kjl}A_{c0j} & 0 & 0\\
 \epsilon^{j}_{\;\;\;kl}B_{mnj} & \frac{1}{2}\delta_{mn}^{ab}\epsilon_{k\;\;\;l}^{\;\;\;j}A_{b0j} & \frac{1}{2}\delta_{mn}^{ab}D_{a\;\;l}^{\;\;k} & 0\\
 \delta_{j}^{i} & 0 & 0 & 0\\
 0 & \delta_{j}^{i} & 0 & 0\\
 0 & 0 & \delta_{b}^{a}\delta_{j}^{i} & 0\\
 0 & 0 & 0 & \delta_{b}^{a}\delta_{j}^{i}\\
 0 & \frac{\Xi}{2}H^{mi\;\;q}_{\;\;\;\;\;j\;\;l} & -\frac{\Xi}{2}E^{ai}_{\;\;\;j} & \frac{\Xi}{2}F^{m\;\;l}_{\;\;\;\;i}\\
 -\frac{\Xi}{2}H^{mi\;\;q}_{\;\;\;\;\;j\;\;l} & \Xi\eta^{abc}\epsilon^{ijk}B_{abk}A_{c0j} & 0 & G^{mjpi} \\
 \frac{\Xi}{2}E^{ai}_{\;\;\;j} & 0 & 0 & 0\\
 \frac{\Xi}{2}F^{m\;\;l}_{\;\;\;\;i} & -G^{kjpi} & 0 & 0
\end{array}
\right)\delta^{3}(x-y),
\label{eq31a}
\end{eqnarray}
where we have defined
\begin{eqnarray}
D_{al}^{\;\;\;i} &\equiv & \delta_{l}^{i}\partial_{a} + \epsilon_{l}^{\;\;ik}\Upsilon_{ak},\nonumber \\
E^{ai}_{\;\;\;j} &\equiv & \eta^{abc}\epsilon^{i}_{\,\;jp}\epsilon^{plk}A_{b0l}A_{c0k} + \eta^{abc}\epsilon^{l}_{\;\;pj}\epsilon^{i}_{\;\;kl}\Upsilon_{b}^{k}\Upsilon_{c}^{p},\nonumber\\ 
F^{m\;\;l}_{\;\;\;\;i} &\equiv & \eta^{mnc}\epsilon_{ijk}\epsilon^{kpl}A_{c0}^{\;\;\;j}A_{n0p},\nonumber\\
G^{mjpi} &\equiv & \eta^{mnc}\epsilon^{j\;\;p}_{\;\;k}\epsilon_{j}^{\;\;li}A_{n0l}\Upsilon_{c}^{k},\nonumber\\
H^{mi\;\;q}_{\;\;\;\;\;j\;\;l} & \equiv &  \eta^{abc}\delta_{ab}^{mn}\epsilon^{i}_{\;\;jk}\epsilon^{qp}_{\;\;\;l}A_{n0p}\Upsilon_{c}^{k}.\nonumber
\end{eqnarray}
Therefore, from the symplectic tensor (\ref{eq31a}) we can identify the generalized FJ brackets by means of 
\begin{eqnarray}
\{\xi_{i}^{(2)}(x),\xi_{j}^{(2)}(y)\}_{FD}=[f^{(2)}_{ij}(x,y)]^{-1},
\end{eqnarray}
thus, the following generalized brackets arise 
\begin{eqnarray}
\left\lbrace A_{c0i}(x), B_{ab}^{\;\;0j}(y) \right\rbrace_{FJ} &=& \frac{1}{2\Xi}\eta_{abc}\delta_{i}^{j}\delta^{3}(x-y),\\
\left\lbrace\Upsilon_{c}^{i}(x), B_{abj}(y)  \right\rbrace_{FJ} &=& \frac{1}{2\Xi}\eta_{abc}\delta_{j}^{i}\delta^{3}(x-y), 
\end{eqnarray}
where we can  observe that the FJ brackets and the Dirac ones coincide to each other  (see the section below). Furthermore, in FJ framework there are less constraints than in Dirac's framework,  in this sense, the FJ is more economical to perform; we will see this point in more details  the following section. Finally, we carry out the counting of physical degrees of freedom. As we have commented above, in FJ formalism there are not a classification of  constraints, they are at the same level, thus, the counting of physical degrees of freedom is performed as $[DF=$ dynamical variables - independent constraints]. In this manner,  there are 18 canonical variables given by $( A_ {coi}, \Upsilon_{a}^{i})$  and 18 independent first class constraints $( \mathop{\Omega}^{(0)}{_{i}}, \mathop{\Omega}^{(0)}{^{00i}},  \mathop{\Omega}^{(0)}{^{0ai}}, \mathop{\Omega}^{(0)}{^{0a}_{\;\;\;0i}} )$;  for $BF$ theory it is well-knew that the constraints are reducible, the reducibility between the constraints is given by $ \partial_a \mathop{\Omega}^{(0)}{^{0ai}}= \epsilon_{ij}{^{k}}\Upsilon_{ak}\mathop{\Omega}^{(0)}{^{0aj}} + \epsilon_{ij}{^{k}}A_{a0k}  \mathop{\Omega}^{(0)}{^{0a}_{\;\;\;0j}}$ and $\partial_a \mathop{\Omega}^{(0)}{^{0a}_{\;\;\;0i}}= \epsilon{_i}^{ jk} \Upsilon_{ak}  \mathop{\Omega}^{(0)}{^{0a}_{\;\;\;0j}} +  \epsilon{_{i}}{^{k}}{_{j}} A_{a0k}\mathop{\Omega}^{(0)}{^{0aj}} $. Therefore,  the theory is devoid of physical degrees of freedom as expected.    It is important to comment, that   all  results found in this section are not reported in the literature. \\
\section{Hamiltonian analysis}
In this section  a pure Dirac's canonical  analysis for the four-dimensional  $BF$  theory  will be  performed, we will follow all  Dirac's  steps in order to obtain the better canonical description of the theory \cite{14a}. For this aim, we start with the Lagrangian given in (\ref{2a})
\begin{eqnarray}
L &=& \Xi\int \eta^{abc}\left[ B_{ab}^{\;\;\;0i}\dot{A}_{c0i} + \frac{1}{2}B_{ab}^{\;\;\;ij}\dot{A}_{cij} \right.\\
&& + \frac{1}{2}A_{0ij}\left( \partial_{c}B_{ab}^{\;\;\;ij} + B_{ab}^{\,\;\;il}A_{c\;\;l}^{\;\;j} + 2B_{ab}^{\,\;\;0i}A_{c0}^{\;\;\;j}\right)\nonumber\\
&& + A_{00i}\left( \partial_{c}B_{ab}^{\,\;\;0i} + B_{ab}^{\;\;\;0j}A_{c\;\;j}^{\;\;i} + B_{ab}^{\;\;\;ij}A_{c0j} \right) \nonumber\\
&& + B_{0a}^{\;\;\;0i}\left( \partial_{b}A_{c0i} - \partial_{c}A_{b0i} + A_{b0j}A_{c\;\;i}^{\;\;j} + A_{c0}^{\;\;\;j}A_{bij} \right) \nonumber\\
&&\left.  + \frac{1}{2}B_{0a}^{\;\;\;ij}\left( \partial_{b}A_{cij} - \partial_{c}A_{bij} + A_{bil}A_{c\;\;j}^{\;\;l} + A_{bi0}A_{c\;\;j}^{\;\;0} - A_{bjl}A_{c\;\;i}^{\;\;l} - A_{bj0}A_{c\;\;i}^{\;\;0} \right)\right] d^{3}x,\nonumber
\end{eqnarray}
by considering the following change of variables  \cite{25, 27}
\begin{eqnarray}
A_{aij} &\equiv & -\epsilon_{ijk}A_{a}^{\;\;k}, \nonumber \\
A_{0ij} &\equiv & -\epsilon_{ijk}A_{0}^{\;\;k},\nonumber \\
B_{abij} &\equiv & -\epsilon_{ijk}B_{ab}^{\;\;\;k},\nonumber \\
B_{0aij} &\equiv & -\epsilon_{ijk}B_{0a}^{\;\;\;k},\nonumber \\
A_{a0i} &\equiv & A_{a0i},\nonumber \\
A_{ai} &\equiv & \Upsilon_{ai},\nonumber \\
A_{0}^{\;\;i} &\equiv & -T^{i},\nonumber \\
A_{00i} &\equiv & -\Lambda_{i},\nonumber \\
B_{0a}^{\;\;\;0i} &\equiv & -\frac{1}{2}\varsigma_{a}^{\;\;i},\nonumber \\
B_{0ai} &\equiv & -\frac{1}{2}\chi_{ai}, 
\end{eqnarray}
the Lagrangian takes the following form
\begin{eqnarray}
L &=& L\left[ A_{a0i}, \Upsilon_{ai}, T_{i}, \Lambda_{i}, \varsigma_{ai}, \chi_{ai}, B_{ab0i}, B_{abi} \right] \nonumber  \\
&=& \int\left[ \Xi\eta^{abc}B_{ab}^{\;\;\;0i}\dot{A}_{c0i} + \Xi\eta^{abc}B_{abi}\dot{\Upsilon}_{c}^{\;\;i} \right. \nonumber\\
&& - T^{i}\left( \partial_{c}\left( \Xi\eta^{abc}B_{abi}\right) + \Xi\eta^{abc}\epsilon^{j}_{\;\;ik}B_{abj}\Upsilon_{c}^{\;\;k} - \Xi\eta^{abc}\epsilon_{jki}B_{ab}^{\;\;\;0j}A_{c0}^{\;\;\;k}  \right)\nonumber\\
&& - \Lambda_{i} \left( \partial_{c}\left( \Xi\eta^{abc}B_{ab}^{\;\;\;0i} \right) -\Xi\eta^{abc}\epsilon^{i}_{\;\;jk}B_{ab}^{\;\;\;0j}\Upsilon_{c}^{\;\;k} - \Xi\eta^{abc}\epsilon^{ijk}B_{abk}A_{c0j}  \right)\nonumber\\
&& -\frac{1}{2} \Xi\eta^{abc}\varsigma_{a}^{\;\;i}\left( \partial_{b}A_{c0i} - \partial_{c}A_{b0i} + \epsilon_{i}^{\;\;jk}A_{b0j}\Upsilon_{ck} - \epsilon_{ijk}A_{c0}^{\;\;\;j}\Upsilon_{b}^{\;\;k} \right) \nonumber\\
&&\left.  - \frac{1}{2} \Xi\eta^{abc}\chi_{ai}\left( \partial_{b}\Upsilon_{c}^{\;\;i} - \partial_{c}\Upsilon_{b}^{\;\;i} + \epsilon^{i}_{\;\;jk}\Upsilon_{b}^{\;\;j}\Upsilon_{c}^{\;\;k} - \epsilon^{ijk}A_{b0j}A_{c0k} \right)\right] d^{3}x.
\end{eqnarray}
In this manner, the canonically momenta of the dynamical variables are given by 
\begin{eqnarray}
p^{a0i} &\equiv & \dfrac{\partial L}{\partial \dot{A}_{a0i}} = \Xi\eta^{abc}B_{bc}^{\;\;\;0i}, \nonumber \\
\pi^{ai} &\equiv & \dfrac{\partial L}{\partial \dot{\Upsilon}_{ai}} = \Xi\eta^{abc}B_{bc}^{\;\;\;i}, \nonumber \\
\hat{T}^{i} &\equiv & \dfrac{\partial L}{\partial \dot{T}_{i}} = 0, \nonumber \\
\hat{\Lambda}^{i} &\equiv & \dfrac{\partial L}{\partial \dot{\Lambda}_{i}} = 0, \nonumber \\
\hat{\varsigma}^{ai} &\equiv & \dfrac{\partial L}{\partial \dot{\varsigma}_{ai}} = 0, \nonumber \\
\hat{\chi}^{ai} &\equiv & \dfrac{\partial L}{\partial \dot{\chi}_{ai}} = 0, \nonumber \\
p^{ab0i} &\equiv & \dfrac{\partial L}{\partial \dot{B}_{ab0i}} = 0, \nonumber  \\
p^{abi} &\equiv & \dfrac{\partial L}{\partial \dot{B}_{abi}} = 0, 
\end{eqnarray}
with the following non-vanishing fundamental Poisson brackets between the fields 
\begin{eqnarray}
\left\lbrace \Upsilon_{ai}(x), \pi^{bj}(y) \right\rbrace &=& \frac{1}{2}\delta_{a}^{b}\delta_{i}^{j}\delta^{3}(x-y),\nonumber \\
\left\lbrace A_{a0i}(x), p^{b0j}(y) \right\rbrace &=& \frac{1}{2}\delta_{a}^{b}\delta_{i}^{j}\delta^{3}(x-y),\nonumber \\
\left\lbrace T_{i}(x), \hat{T}^{j}(y) \right\rbrace &=& \frac{1}{2}\delta_{i}^{j}\delta^{3}(x-y),\nonumber \\
\left\lbrace \Lambda_{i}(x), \hat{\Lambda}^{j}(y) \right\rbrace &=& \frac{1}{2}\delta_{i}^{j}\delta^{3}(x-y),\nonumber \\
\left\lbrace \varsigma_{ai}(x), \hat{\varsigma}^{bj}(y) \right\rbrace &=& \delta_{a}^{b}\delta_{i}^{j}\delta^{3}(x-y),\nonumber \\
\left\lbrace \chi_{ai}(x), \hat{\chi}^{bj}(y) \right\rbrace &=& \delta_{a}^{b}\delta_{i}^{j}\delta^{3}(x-y),\nonumber \\
\left\lbrace B_{ab0i}(x), p^{de0j}(y) \right\rbrace &=& \frac{1}{4}\left( \delta_{a}^{d}\delta_{b}^{e} - \delta_{a}^{e }\delta_{b}^{d}\right) \delta_{i}^{j}\delta^{3}(x-y),\nonumber \\
\left\lbrace B_{abi}(x), p^{dej}(y) \right\rbrace &=& \frac{1}{4}\left( \delta_{a}^{d}\delta_{b}^{e} - \delta_{a}^{e }\delta_{b}^{d}\right) \delta_{i}^{j}\delta^{3}(x-y). 
\end{eqnarray}
Furthermore, from the definition of the momenta, we identify the following 60 primary constraints 
\begin{eqnarray}
\phi_{1}^{a0i} &\equiv & p^{a0i} - \Xi\eta^{abc}B_{bc}^{\;\;\;0i} \approx 0, \nonumber \\
\phi_{2}^{ai} &\equiv & \pi^{ai} - \Xi\eta^{abc}B_{bc}^{\;\;\;i} \approx 0,\nonumber \\
\phi_{3}^{i} &\equiv & \hat{T}^{i} \approx 0,\nonumber \\
\phi_{4}^{i} &\equiv & \hat{\Lambda}^{i} \approx 0,\nonumber \\
\phi_{5}^{ai} &\equiv & \hat{\varsigma}^{ai} \approx 0,\nonumber \\
\phi_{6}^{ai} &\equiv & \hat{\chi}^{ai} \approx 0,\nonumber \\
\phi_{7}^{ab0i} &\equiv & p^{ab0i} \approx 0,\nonumber \\
\phi_{8}^{abi} &\equiv & p^{abi} \approx 0.
\end{eqnarray}
The canonical Hamiltonian  of the theory is given by 
\begin{eqnarray}
H_{c} &=& \int\left[\dot{A}_{a0i}p^{a0i} + \dot{\Upsilon}_{ai}\pi^{ai} + \dot{T}_{i}\hat{T}^{i} + \dot{\Lambda}_{i}\hat{\Lambda}^{i} +  \dot{\varsigma}_{ai}\hat{\varsigma}^{ai} + \dot{B}_{ab0i}p^{ab0i} + \dot{B}_{abi}p^{abi} - L \right] d^{3}x\nonumber\\
&=& \int\left[  T^{i}\left( \partial_{a}\pi^{a}_{\;\;i} - \epsilon^{\;\;jk}_{i}\pi^{a}_{\;\;j}\Upsilon_{ak} - \epsilon_{ijk}p^{a0j}A_{a0}^{\;\;\;k}  \right)\right.\nonumber\\
&& + \Lambda_{i} \left( \partial_{a}p^{a0i} - \epsilon^{i}_{\;\;jk}p^{a0j}\Upsilon_{a}^{\;\;k} - \epsilon^{ijk}\pi^{a}_{\;\;k}A_{a0j}  \right)\nonumber\\
&& + \frac{1}{2} \Xi\eta^{abc}\varsigma_{a}^{\;\;i}\left( \partial_{b}A_{c0i} - \partial_{c}A_{b0i} + \epsilon_{i}^{\;\;jk}A_{b0j}\Upsilon_{ck} + \epsilon_{ijk}A_{c0}^{\;\;\;j}\Upsilon_{b}^{\;\;k} \right) \nonumber\\
&&\left.  + \frac{1}{2} \Xi\eta^{abc}\chi_{ai}\left( \partial_{b}\Upsilon_{c}^{\;\;i} - \partial_{c}\Upsilon_{b}^{\;\;i} + \epsilon^{i}_{\;\;jk}\Upsilon_{b}^{\;\;j}\Upsilon_{c}^{\;\;k} - \epsilon^{ijk}A_{b0j}A_{c0k} \right)\right] d^{3}x, 
\end{eqnarray}
 by adding the primary constraints we obtain  the primary Hamiltonian 
\begin{eqnarray}
H_{1} &=& \int\left[  H_c + T^{i}\left( \partial_{a}\pi^{a}_{\;\;i} - \epsilon^{\;\;jk}_{i}\pi^{a}_{\;\;j}\Upsilon_{ak} - \epsilon_{ijk}p^{a0j}A_{a0}^{\;\;\;k}  \right)\right.\nonumber\\
&& + \Lambda_{i} \left( \partial_{a}p^{a0i} - \epsilon^{i}_{\;\;jk}p^{a0j}\Upsilon_{a}^{\;\;k} - \epsilon^{ijk}\pi^{a}_{\;\;k}A_{a0j}  \right)\nonumber\\
&& + \frac{1}{2} \Xi\eta^{abc}\varsigma_{a}^{\;\;i}\left( \partial_{b}A_{c0i} - \partial_{c}A_{b0i} + \epsilon_{i}^{\;\;jk}A_{b0j}\Upsilon_{ck} + \epsilon_{ijk}A_{c0}^{\;\;\;j}\Upsilon_{b}^{\;\;k} \right) \nonumber\\
&&  + \frac{1}{2} \Xi\eta^{abc}\chi_{ai}\left( \partial_{b}\Upsilon_{c}^{\;\;i} - \partial_{c}\Upsilon_{b}^{\;\;i} + \epsilon^{i}_{\;\;jk}\Upsilon_{b}^{\;\;j}\Upsilon_{c}^{\;\;k} - \epsilon^{ijk}A_{b0j}A_{c0k} \right)\nonumber\\
&& + \lambda_{a0i}\left( p^{a0i} - \Xi\eta^{abc}B_{bc}^{\,\;\;0i} \right)  + \lambda_{ai}\left( \pi^{ai} - \Xi\eta^{abc}B_{bc}^{\;\;\;i} \right) + \alpha_{i}\hat{T}^{i} + \beta_{i}\hat{\Lambda}^{i}\nonumber\\
&& \left.  + \theta_{ai}\hat{\varsigma}^{ai} + \mu_{ai}\hat{\chi}^{ai} + \lambda_{ab0i}p^{ab0i} + \lambda_{abi}p^{abi}\right] d^{3}x, 
\end{eqnarray}
where $\lambda_{a0i}, \lambda_{ai}, \alpha_{i}, \beta_{i}, \theta_{ai}, \mu_{ai}, \lambda_{ab0i}, \lambda_{abi}$ are Lagrange multipliers enforcing the primary constraints. From consistency of the constraints, we identify  the following 24 secondary constraints 
\begin{eqnarray}
\dot{\phi}_{3}^{i} \approx 0 \Rightarrow \varphi_{3}^{i} &\equiv & -\frac{1}{2}\left[ \partial_{a}\pi^{ai} - \epsilon^{ijk}\pi^{a}_{\;\;j}\Upsilon_{ak} - \epsilon^{i}_{\;\;jk}p^{a0j}A_{a0}^{\;\;\;k} \right] \approx 0,\\
\dot{\phi}_{4}^{i} \approx 0 \Rightarrow \varphi_{4}^{i} &\equiv & -\frac{1}{2}\left[ \partial_{a}p^{a0i} - \epsilon^{i}_{\;\;jk}p^{a0j}\Upsilon_{a}^{\;\;k} - \epsilon^{ijk}\pi^{a}_{\;\;k}A_{a0j} \right] \approx 0,\\
\dot{\phi}_{5}^{ai} \approx 0 \Rightarrow \varphi_{5}^{ai} &\equiv & -\frac{\Xi}{2}\eta^{abc}\left( \partial_{b}A_{c}^{\;\;0i} - \partial_{c}A_{b}^{\;\;0i} - \epsilon^{ijk}A_{b0j}\Upsilon_{ck} + \epsilon^{ijk}A_{c0j}\Upsilon_{bk} \right)\nonumber\\
&& \approx 0,\\
\dot{\phi}_{6}^{ai} \approx 0 \Rightarrow \varphi_{6}^{ai} &\equiv & -\frac{\Xi}{2}\eta^{abc}\left[ \partial_{b}\Upsilon_{c}^{\;\;i} - \partial_{c}\Upsilon_{b}^{\;\;i} + \epsilon^{ijk}\Upsilon_{bj}\Upsilon_{ck} - \epsilon^{ijk}A_{b0j}A_{c0k} \right]\nonumber\\
&& \approx 0, 
\end{eqnarray}
and   the  following  36 Lagrange multipliers
\begin{eqnarray}
\lambda_{ab}^{\;\;\;0i} &\approx & \frac{1}{2\Xi}\left( \eta_{abc}\epsilon^{i}_{\;\;jk}T^{j}p^{c0k} - \eta_{abc}\epsilon^{ijk}\Lambda_{j}\pi^{c}_{\;\;k} + \Xi\partial_{b}\varsigma_{a}^{\;\;i} - \Xi\partial_{a}\varsigma_{b}^{\;\;i} - \Xi\epsilon^{i\;\;k}_{\;\;j}\varsigma_{a}^{\;\;j}\Upsilon_{bk}\right.\nonumber\\
&& \left. + \Xi\epsilon^{i\;\;k}_{\;\;j}\varsigma_{b}^{\,\;j}\Upsilon_{ak} + \epsilon^{ijk}\chi_{aj}A_{b0k} - \epsilon^{ijk}\chi_{bj}A_{a0k}  \right),\\
\lambda_{ab}^{\;\;\;i} &\approx & \frac{1}{2\Xi}\left( \eta_{abc}\epsilon^{i\;\;k}_{\;j}T^{j}\pi^{c}_{\;\;k} + \eta_{abc}\epsilon^{ij}_{\;\;\;k}p^{c0k}\Lambda_{j} - \Xi\epsilon^{i\;\;k}_{\;\;j}\varsigma_{a}^{\;\;j}A_{b0k} + \Xi\epsilon^{i\;\;k}_{\;\;j}\varsigma_{b}^{\;\;j}A_{a0k}\right.\nonumber\\
&& \left. + \Xi\partial_{b}\chi_{a}^{\;\;i} - \Xi\partial_{a}\chi_{b}^{\;\;i} - \Xi\epsilon^{ijk}\chi_{aj}\Upsilon_{bk} + \Xi\epsilon^{ijk}\chi_{bj}\Upsilon_{ak}  \right),\\
\lambda_{a0i} &\approx & 0,\\
\lambda_{ai} &\approx & 0. 
\end{eqnarray}
For this theory there are not tertiary  constraints. Hence, in order to perform  the classification of the constraints  in first class and second class we proceed to calculate the following matrix whose entries are the  Poisson brackets between all the constraints, it is  
\begin{eqnarray}
\left\lbrace \varphi_{3}^{i}(x), \varphi^{l}_{3}(y) \right\rbrace &=& -\frac{1}{4}\epsilon^{ilk}\varphi_{3k}\delta^{3}(x-y) \approx 0,\\
\left\lbrace \varphi_{4}^{i}(x), \varphi_{4}^{l}(y) \right\rbrace &=& \frac{1}{4}\epsilon^{ilk}\varphi_{3k}\delta^{3}(x-y)\approx 0.\\
\left\lbrace \phi_{1}^{a0i}(x), \phi_{7}^{de0l}(y) \right\rbrace  &=& \frac{\Xi}{2}\eta^{ade}\eta^{il}\delta^{3}(x-y),\\
\left\lbrace \phi_{1}^{a0i}(x), \varphi_{3}^{l}(y) \right\rbrace &=& - \frac{1}{4}\epsilon^{il}_{\;\;\;j}p^{a0j}\delta^{3}(x-y),\\
\left\lbrace \phi_{1}^{a0i}(x), \varphi_{4}^{l}(y) \right\rbrace &=& \frac{1}{4}\epsilon^{ilj}\pi^{a}_{\;\;j}\delta^{3}(x-y),\\
\left\lbrace \phi_{1}^{a0i}(x), \varphi_{5}^{dl}(y) \right\rbrace &=& -\frac{\Xi}{2}\left[ -\eta^{il}\eta^{ade}\partial_{x,e} + \eta^{ade}\epsilon^{ilj}\Upsilon_{ej} \right] \delta^{3}(x-y),\\
\left\lbrace \phi_{1}^{a0i}(x), \varphi_{6}^{dl}(y) \right\rbrace &=& -\frac{\Xi}{2}\eta^{ade}\epsilon^{ilj}A_{e0j}\delta^{3}(x-y),\\
\left\lbrace \phi_{2}^{ai}(x), \phi^{del}_{8}(y) \right\rbrace &=& -\frac{\Xi}{2}\eta^{il}\eta^{ade}\delta^{3}(x-y),\\
\left\lbrace \phi_{2}^{ai}(x), \varphi_{3}^{l}(y) \right\rbrace &=& -\frac{1}{4}\epsilon^{ilj}\pi^{a}_{\;\;j}\delta^{3}(x-y),\\
\left\lbrace \phi_{2}^{ai}(x), \varphi_{4}^{l}(y) \right\rbrace &=& -\frac{1}{4}\epsilon^{il}_{\;\;\;j}p^{a0j}\delta^{3}(x-y),\\
\left\lbrace \phi_{2}^{ai}(x), \varphi_{5}^{dl}(y) \right\rbrace &=& -\frac{\Xi}{2}\eta^{ade}\epsilon^{ilj}A_{e0j}\delta^{3}(x-y),\\
\left\lbrace \phi_{2}^{ai}(x), \varphi_{6}^{dl}(y) \right\rbrace &=& \frac{\Xi}{2}\eta^{ade}\left[ -\eta^{il}\partial_{x,e} + \epsilon^{ilj}\Upsilon_{ej} \right] \delta^{3}(x-y),\\
\left\lbrace \varphi_{3}^{i}(x), \varphi_{4}^{l}(y) \right\rbrace &=& - \frac{1}{4}\epsilon^{il}_{\;\;\;j}\varphi_{4}^{j}\delta^{3}(x-y) \approx 0,\\
\left\lbrace \varphi_{3}^{i}(x), \varphi_{5}^{dl}(y) \right\rbrace &=& -\frac{1}{4}\epsilon^{il}_{\;\;\;k}\varphi_{5}^{dk}\delta^{3}(x-y) \approx 0,\\
\left\lbrace \varphi_{3}^{i}(x), \varphi_{6}^{dl}(y) \right\rbrace  &=& -\frac{1}{4}\epsilon^{il}_{\;\;\;k}\varphi_{6}^{dk}\delta^{3}(x-y)\approx 0,\\
\left\lbrace \varphi_{4}^{i}(x), \varphi_{5}^{dl}(y) \right\rbrace &=& \frac{1}{4}\epsilon^{il}_{\;\;\;k}\varphi_{6}^{k}\delta^{3}(x-y) \approx 0,\\
\left\lbrace \varphi_{4}^{i}(x), \varphi_{6}^{dl}(y) \right\rbrace &=& -\frac{1}{4}\epsilon^{il}_{\;\;\;j}\varphi^{dj}_{5}\delta^{3}(x-y) \approx 0.
\end{eqnarray}
we find that this  matrix has rank = 36 and 48 null vectors. From the null vectors we find the following 48 first class constraints  
\begin{eqnarray}
\gamma_{1}^{i} &\equiv & \hat{T}^{i} \approx 0, \nonumber\\
\gamma_{2}^{i} &\equiv & \hat{\Lambda}^{i} \approx 0,\nonumber \\
\gamma_{3}^{ai} &\equiv & \hat{\varsigma}^{ai} \approx 0,\nonumber\\
\gamma_{4}^{ai} &\equiv & \hat{\chi}^{ai} \approx 0,\nonumber \\
\gamma_{5\;i} &\equiv & \partial_{a}\pi^{a}_{\;\;i} - \epsilon_{i}^{\;\;jk}\pi^{a}_{\;\;j}\Upsilon_{ak} - \epsilon_{ijk}p^{a0j}A_{a0}^{\;\;\;k}\nonumber\\
&& + \frac{1}{2\Xi}\eta_{abc}\epsilon_{ijk}\left( \pi^{aj}p^{bck} - p^{a0j}p^{bc0k} \right)\approx 0,\nonumber \\
\gamma_{6}^{0i} &\equiv & \partial_{a}p^{a0i} - \epsilon^{i}_{\;\;jk}p^{a0j}\Upsilon_{a}^{\;\;k} - \epsilon^{ijk}\pi^{a}_{\;\;k}A_{a0j}\nonumber\\
&& + \frac{1}{2\Xi}\eta_{abc}\epsilon^{i}_{\;\;jk}\left( \pi^{aj}p^{bc0k} + p^{a0j}p^{bck} \right) \approx 0, \nonumber \\
\gamma_{7\;0i}^{a} &\equiv & \frac{\Xi}{2}\eta^{abc}\left( \partial_{b}A_{c0i} - \partial_{c}A_{b0i} + \epsilon_{i}^{\;\;jk}A_{b0j}\Upsilon_{ck} - \epsilon_{ijk}A_{c0}^{\,\;\;j}\Upsilon_{b}^{\;\;k}\right) \nonumber\\
&& - \partial_{b}p^{ab}_{\;\;\;0i} + \epsilon_{i}^{\;\;jk}\Upsilon_{bk}p^{ab}_{\;\;\;0j} + \epsilon_{i}^{\;\;jk}A_{b0k}p^{ab}_{\;\;\;j} \approx 0, \nonumber \\
\gamma_{8}^{ai} &\equiv & \frac{\Xi}{2}\eta^{abc}\left[ \partial_{b}\Upsilon_{c}^{\;\;i} - \partial_{c}\Upsilon_{b}^{\;\;i} + \epsilon^{ijk}\Upsilon_{bj}\Upsilon_{ck} - \epsilon^{ijk}A_{b0j}A_{c0k} \right]\nonumber\\
&& - \partial_{b}p^{abi} - \epsilon^{ijk}A_{b0k}p^{ab}_{\;\;\;0j} + \epsilon^{ijk}\Upsilon_{bk}p^{ab}_{\,\;\;j} \approx 0.
\label{first}
\end{eqnarray}
and the rank allows  us identify the following 36 second class constraints 
\begin{eqnarray}
\Gamma_{1}^{a0i} &\equiv & p^{a0i} - \Xi\eta^{abc}B_{bc}^{\;\;\;0i} \approx 0,\nonumber \\
\Gamma_{2}^{ai} &\equiv & \pi^{ai} - \Xi\eta^{abc}B_{bc}^{\;\;\;i} \approx 0,\nonumber \\
\Gamma_{3}^{ab0i} &\equiv & p^{ab0i} \approx 0, \nonumber\\
\Gamma_{4}^{abi} &\equiv & p^{abi} \approx 0.
\label{second}
\end{eqnarray}
It is important to remark  that the complete structure of the constraints (\ref{first}) is not reported  in the literature and this is a result  of performing  a pure Dirac's formulation. In fact, by working with  the standard form, it is not possible to obtain a full   structure of the constraints and the algebra could not be closed just as is present in four-dimensional Palatini's theory \cite{14}. Furthermore,    in order to compare the symplectic framework with the Dirac one, it is necessary to follow all  steps of  the Dirac formulation as has been developed in this paper. \\
With  all   information obtained, we can carryout   the counting of physical degrees of the theory in the following form; there are 120 dynamical variables, 48 first class constraints and 36 second class constraints, hence we obtain -6 degrees of freedom. However, it is well-known that $BF$ theory is a reducible theory,  this is, the constraints are not independent to each other. The reducibility of the constraints are given by 
\begin{eqnarray}
\partial_{a}\gamma^{a}_{7\;0i} &=& \epsilon_{i}^{\;\;jk}\Upsilon_{ak}\gamma^{a}_{7\;0j} + \epsilon_{i}^{\;\;jk}A_{a0k}\gamma_{8\;j}^{a} + \frac{1}{2}\epsilon_{i}^{\;\;jk}F_{abk}\Gamma_{3\;\;0j}^{ab} + \frac{1}{2}\epsilon_{i}^{\;\;jk}F_{ab0k} \Gamma_{4\;\;j}^{ab}\nonumber\\
\partial_{a}\gamma_{8}^{ai} &=&  \epsilon^{ijk}\Upsilon_{ak}\gamma_{8\;j}^{a} - \epsilon^{ijk}A_{a0k}\gamma_{7\;0j}^{a} - \frac{1}{2}\epsilon^{ijk}F_{ab0k}\Gamma^{ab}_{3\;\;0j} + \frac{1}{2}\epsilon^{ijk}F_{abk}\Gamma_{4\;\;j}^{ab} ,\nonumber
\end{eqnarray}
hence, there are  42 independent first class constraints. Therefore, by performing the counting of physical degrees of freedom we conclude that the theory is devoid of degrees of freedom, the theory is a topological one as expected. \\
The algebra between the constraints is given by  
\begin{eqnarray}
\left\lbrace \Gamma_{1}^{a0i}(x), \Gamma_{3}^{de0l}(y) \right\rbrace &=& \frac{\Xi}{2}\eta^{ade}\eta^{il}\delta^{3}(x-y), \nonumber \\
\left\lbrace \Gamma_{2}^{ai}(x), \Gamma_{4}^{del}(y) \right\rbrace &=& -\frac{\Xi}{2}\eta^{ade}\eta^{il}\delta^{3}(x-y), \nonumber \\
\left\lbrace \gamma_{5}^{i}(x), \gamma_{5}^{l}(y) \right\rbrace &=& \frac{1}{2}\epsilon^{il}_{\;\;\;j}\gamma_{5}^{j}\delta^{3}(x-y)\approx 0, \nonumber\\
\left\lbrace \gamma_{5}^{i}(x),\gamma^{0l}_{6}(y) \right\rbrace &=& \frac{1}{2}\epsilon^{il}_{\,\;\;j}\gamma_{6}^{j}\delta^{3}(x-y) \approx 0,\nonumber \\
\left\lbrace \gamma_{5\;i}(x), \gamma_{7\; 0l}^{d}(y) \right\rbrace &=& \frac{1}{2}\epsilon_{il}^{\,\;\;k}\gamma_{7\;0k}^{d} - \frac{1}{4\Xi}\eta_{abc}\Gamma^{bc}_{4\;\;l}\Gamma^{da}_{3\,\;\;0i} + \frac{1}{4\Xi}\eta_{abc}\Gamma^{bc0}_{3\;\;l}\Gamma^{da}_{4\;\;i} \nonumber\\
&\approx &0, \nonumber\\
\left\lbrace \gamma_{5\; i}(x), \gamma_{8}^{dl}(y) \right\rbrace &=& \frac{1}{2}\epsilon_{i}^{\;\;lk}\gamma_{8\;k}^{d} - \frac{1}{4\Xi}\eta_{abc}\Gamma_{3}^{bc0l}\Gamma_{3\,\;\;0i}^{da} + \frac{1}{4\Xi}\delta_{i}^{l}\eta_{abc}\Gamma_{3}^{bc0k}\Gamma_{3\,\;\;0k}^{da}\nonumber\\
&& + \frac{1}{4\Xi}\eta_{abc}\delta_{i}^{l}\Gamma_{4}^{bck}\Gamma_{4\,\;\;k}^{da} - \frac{1}{4\Xi}\eta_{abc}\Gamma_{4}^{bcl}\Gamma_{4\,\,\;i}^{da}\nonumber\\
&\approx & 0, \nonumber \\ 
\left\lbrace \gamma_{6}^{0i}(x), \gamma_{6}^{0l}(y) \right\rbrace &=& - \frac{1}{2}\epsilon^{ilj}\gamma_{5\;j}\delta^{3}(x-y)\approx 0, \nonumber\\
\left\lbrace \gamma_{6}^{0i}(x), \gamma_{7\;0l}^{d}(y) \right\rbrace &=& \frac{1}{2}\epsilon_{\,\;l}^{i\;\;k}\gamma_{8k}^{d} - \frac{1}{4\Xi}\delta_{i}^{l}\eta_{abc}\Gamma_{3\;\;\;k}^{bc0}\Gamma_{3}^{da0k} + \frac{1}{4\Xi}\eta_{abc}\Gamma_{3\;\;\;l}^{bc0}\Gamma_{3}^{da0i}\nonumber\\
&& + \frac{1}{4\Xi}\eta_{abc}\delta^{i}_{l}\Gamma_{4\;\;\;k}^{bc}\Gamma_{4}^{dak} - \frac{1}{4\Xi}\eta_{abc}\Gamma_{4\;\;\;l}^{bc}\Gamma_{4}^{dai}\nonumber\\
&\approx &0, \nonumber\\
\left\lbrace \gamma_{6}^{0i}(x), \gamma_{8}^{dl}(y) \right\rbrace &=& -\frac{1}{2}\epsilon {ilj}\gamma_{7\; 0j}^{d} + \frac{1}{4\Xi}\eta_{abc}g^{il}\Gamma_{3}^{bc0k}\Gamma^{da}_{4\;\,\;k} - \frac{1}{4\Xi}\eta_{abc}\Gamma_{3}^{bc0l}\Gamma_{4}^{dai}\nonumber\\
&& - \frac{1}{4\Xi}\eta_{abc}\Gamma_{4}^{bcl}\Gamma_{3}^{da0i} + \frac{1}{4\Xi}g^{il}\eta_{abc}\Gamma_{4}^{bck}\Gamma_{3\;\,\;k}^{da0}\nonumber\\
&\approx &0, \nonumber \\
\left\lbrace \gamma_{7\; 0i}^{a}(x), \Gamma_{1}^{d0l}(y) \right\rbrace &=& -\frac{1}{2}\epsilon_{i}^{\;\;lj}p^{ad}_{\;\;\;j}\delta^{3}(x-y)\nonumber\\
&=& -\frac{1}{2}\epsilon_{i}^{\;\;lj}\Gamma^{ad}_{4\;\;j}\delta^{3}(x-y)\approx 0, \nonumber \\
\left\lbrace \gamma_{7\; 0i}^{a}(x), \Gamma_{2}^{dl}(y) \right\rbrace &=& -\frac{1}{2}\epsilon_{i}^{\;\;lj}p^{ad}_{\;\;\;0j}\delta^{3}(x-y)\nonumber\\
&=& -\frac{1}{2}\epsilon_{i}^{\;\;lj}\Gamma^{ad}_{3\;\;0j}\delta^{3}(x-y)\approx 0, \nonumber \\
\left\lbrace \gamma_{8}^{ai}(x), \Gamma^{d0l}_{1}(y) \right\rbrace &=& \frac{1}{2}\epsilon^{ilj}p^{ad}_{\,\;\;0j}\nonumber\\
&=& \frac{1}{2}\epsilon^{ilj}\Gamma^{ad}_{3\;\;0j}\approx 0, \nonumber \\
\left\lbrace \gamma_{8}^{ai}(x), \Gamma^{dl}_{2}(y) \right\rbrace &=& \frac{1}{2}\epsilon^{ilj}p^{ad}_{\,\;\;j}\nonumber\\
&=& \frac{1}{2}\epsilon^{ilj}\Gamma^{ad}_{4\;\;j}\approx 0,
\end{eqnarray}
where we observe that the algebra is closed and it   obeys the rules of the canonical formalism   of  the  algebra between the constraints \cite{14a, 15, 19}, namely, the result    of the Poisson brackets between first class constraints with first class  must be linear in first class constraints and square  in second class;  the result  of the Poisson brackets between first class constraints with second class constraints must be linear in first class constraints and linear  in second class constraints. We can observe that our results are in full agreement  with these rules. Furthermore,  the constraints     $\gamma_{5\; i}$  and $  \gamma_{6}^{0i} $ are identified as generators of rotations and boost respectively whereas 
$\gamma_{7}{^{0i}}$ and  $\gamma_{8}{^{ai}}$ are generators of translations, this can be seen from the algebra between these constraints.  \\
On the other hand,  first class constraints are generators of gauge transformations. Hence, by defining the  gauge generator in terms of first class constraints 
\begin{eqnarray}
G &=& \int \left[ \theta_{1,i}\gamma_{1}^{i} + \theta_{2,i}\gamma^{i}_{2} + \theta_{3,ai}\gamma_{3}^{ai} + \theta_{4,ai}\gamma^{ai}_{4} + \theta_{5}^{i}\gamma_{5,i} + \theta_{6,0i}\gamma_{6}^{0i} + \theta_{7,a}^{0i}\gamma_{7,0i}^{a} + \theta_{8,ai}\gamma_{8}^{ai} \right] d^{3}x, \nonumber
\end{eqnarray}
we find the following gauge transformations
\begin{eqnarray}
\delta A_{d0l} &=& \left\lbrace A_{d0l}, G \right\rbrace \\
&=& \frac{1}{2}\left[ -\partial_{d}\theta_{6,0l} + \theta_{5}^{i}\epsilon_{lik}A_{d0}^{\;\;\;k} - \epsilon^{i}_{\;lk}\theta_{6,0i}\Upsilon_{d}^{\;k} + \frac{1}{2\Xi}\eta_{dbc}\epsilon_{lik}\theta_{5}^{i}p^{bc0k} + \frac{1}{2\Xi}\eta_{dbc}\epsilon^{i}_{\;\;lk}p^{bck} \right] \nonumber\\
\delta \Upsilon_{dl} &=& \left\lbrace \Upsilon_{dl}, G \right\rbrace \\
&=& \frac{1}{2}\left[ -\partial_{d}\theta_{5,l} + \epsilon_{li}^{\;\;\;k}\theta_{5}^{i}\Upsilon_{dk} - \epsilon_{l}^{\;\;ij}\theta_{6,0i}A_{d0l} - \frac{1}{2\Xi}\eta_{dbc}\epsilon_{lik}\theta_{5}^{i}p^{bck} + \frac{1}{2\Xi}\eta_{dbc}\epsilon^{i}_{\;\;lk}\theta_{6,0i}p^{bc0k} \right] \nonumber\\
\delta T_{l} &=& \left\lbrace T_{l}, G \right\rbrace = \theta_{1,l}\\
\delta \Lambda_{i} &=& \left\lbrace \Lambda_{l}, G \right\rbrace = \theta_{2,l}\\
\delta \varsigma_{dl} &=& \left\lbrace \varsigma_{dl}, G \right\rbrace = \theta_{3,dl} \\
\delta \chi_{dl} &=& \left\lbrace \chi_{dl}, G \right\rbrace = \theta_{4,dl}\\
\delta B_{de0l} &=& \left\lbrace B_{de0l}, G \right\rbrace \nonumber\\
&=& \frac{1}{4}\left[ \partial_{e}\theta_{7,d0l} - \partial_{d}\theta_{7,e0l} - \theta_{l}^{\;\;ij}\theta_{7,d0j}\Upsilon_{ek} + \epsilon_{l}^{\;\;ij}\theta_{7,e0i}\Upsilon_{dj} - \epsilon_{l}^{\;\;ik}\theta_{8,di}A_{e0k} + \epsilon_{l}^{\;\;ik}\theta_{8,ei}A_{dok}\right.\nonumber\\
&& \left. - \frac{1}{\Xi}\eta_{dea}\epsilon_{lij}\theta_{5}^{i}p^{a0j} + \frac{1}{\Xi}\eta_{dea}\epsilon_{l\;\;j}^{\;\;i}\theta_{6,0i}\pi^{aj}  \right] \\
\delta B_{del} &=& \left\lbrace B_{del}, G \right\rbrace \nonumber\\
&=& \frac{1}{4}\left[ \partial_{e}\theta_{8,dl} - \partial_{d}\theta_{8,el} - \epsilon_{l}^{\;\;ik}\theta_{8,di}\Upsilon_{ek} + \epsilon_{l}^{\;\;ik}\theta_{8,ei}\Upsilon_{dk} - \epsilon_{li}^{\;\;\;k}\theta_{7,d}^{0i}A_{e0k} + \epsilon_{li}^{\;\;\;k}\theta_{7,e}^{0i}A_{d0k}\right.\nonumber\\
&&\left. + \frac{1}{\Xi}\eta_{dea}\epsilon_{lij}\theta_{5}^{i}\pi^{aj} + \frac{1}{\Xi}\eta_{dea}\epsilon_{l\;\;j}^{\;\;i}\theta_{6,0i}p^{a0j}  \right] \\
\delta p^{d0l} &=& \left\lbrace p^{d0l}, G \right\rbrace \nonumber\\
&=& \frac{1}{2}\left[ \Xi\eta^{dab}\partial_{b}\theta_{7,a}^{0l} + \epsilon^{l}_{\;\;ij}\theta_{5}^{i}p^{d0j} - \epsilon^{lij}\theta_{6,0i}\pi^{d}_{\,\;k} - \epsilon^{l\;\;j}_{\;\;j}\theta_{7,a}^{\;\;\;0i}p^{ad}_{\;\;\;j} + \epsilon^{lij}\theta_{8,ai}p^{ad}_{\,\;\;0j}\right.\nonumber\\
&& \left. - \Xi\eta^{dac}\epsilon^{l\;\;k}_{\,\;i}\theta_{7,a}^{\;\;0i}\Upsilon_{ck} + \Xi\eta^{dac}\epsilon^{lik}\theta_{8, ai}A_{c0k}  \right]\\ 
\delta \pi^{dl} &=& \left\lbrace \pi^{dl}, G \right\rbrace \nonumber\\
&=& \frac{1}{2}\left[ \Xi\eta^{dab}\partial_{b}\theta_{8,a}^{l} + \epsilon^{l\;\;j}_{\;\;i}\theta_{5}^{i}\pi^{d}_{\;\;j} + \epsilon^{li}_{\;\;\;j}\theta_{6,0i}p^{d0j} + \epsilon^{l\;\;j}_{\;\;i}\theta_{7,a}^{\;\;0i}p^{da}_{\;\;\;0j} + \epsilon^{lij}\theta_{8,ai}p^{da}_{\,\;\;j} \right.\nonumber\\
&& \left.  - \Xi\eta^{dab}\epsilon^{l\;\;j}_{\;\;i}\theta_{7,a}^{\;\;0i}A_{b0j} - \Xi\eta^{dac}\epsilon^{lik}\theta_{8,ai}\Upsilon_{ck} \right] \\
\delta \hat{T}^{l} &=& \left\lbrace \hat{T}^{l},G \right\rbrace = 0,\\
\delta \hat{\Lambda}^{l} &=& \left\lbrace \hat{\Lambda}^{l},G \right\rbrace = 0,\\
\delta \hat{\varsigma}^{dl} &=& \left\lbrace \hat{\varsigma}^{dl},G \right\rbrace = 0,\\
\delta \hat{\chi}^{dl} &=& \left\lbrace \hat{\chi}^{l},G \right\rbrace = 0,\\
\delta p^{de0l} &=& \left\lbrace p^{de0l},G \right\rbrace = 0,\\
\delta p^{del} &=& \left\lbrace p^{del},G \right\rbrace = 0.
\end{eqnarray}

Now, with the second class  constraints we  can calculate  the Dirac brackets. In order to perform this aim  we calculate  the matrix $C^{\alpha\beta} = \left\lbrace \Gamma^{\alpha}, \Gamma^{\beta} \right\rbrace $ whose entries are given by the Poisson brackets of the second class constraints, namely  
\begin{eqnarray}
C_{\alpha \beta} &=& \left(
\begin{array}{cccc}
\left\lbrace \Gamma_{1}^{a0i}, \Gamma_{1}^{d0l} \right\rbrace  & \left\lbrace \Gamma_{1}^{a0i}, \Gamma_{2}^{dl} \right\rbrace & \left\lbrace \Gamma_{1}^{a0i}, \Gamma_{3}^{de0l} \right\rbrace & \left\lbrace \Gamma_{1}^{a0i}, \Gamma_{4}^{del} \right\rbrace \\
\left\lbrace \Gamma_{2}^{ai}, \Gamma_{1}^{d0l} \right\rbrace & \left\lbrace \Gamma_{2}^{ai}, \Gamma_{2}^{dl} \right\rbrace & \left\lbrace \Gamma_{2}^{ai}, \Gamma_{3}^{de0l} \right\rbrace & \left\lbrace \Gamma_{2}^{ai}, \Gamma_{4}^{del} \right\rbrace \\
\left\lbrace \Gamma_{3}^{ab0i}, \Gamma_{1}^{d0l} \right\rbrace & \left\lbrace \Gamma_{3}^{ab0i}, \Gamma_{2}^{dl} \right\rbrace & \left\lbrace \Gamma_{3}^{ab0i}, \Gamma_{3}^{de0l} \right\rbrace & \left\lbrace \Gamma_{3}^{ab0i}, \Gamma_{4}^{del} \right\rbrace \\
\left\lbrace \Gamma_{4}^{abi}, \Gamma_{1}^{d0l} \right\rbrace & \left\lbrace \Gamma_{4}^{abi}, \Gamma_{2}^{dl} \right\rbrace & \left\lbrace \Gamma_{4}^{abi}, \Gamma_{3}^{de0l} \right\rbrace & \left\lbrace \Gamma_{4}^{abi}, \Gamma_{4}^{del} \right\rbrace
\end{array}
\right)\nonumber\\
&=& \left(
\begin{array}{cccc}
0  & 0 & \frac{\Xi}{2}\eta^{ade}\eta^{il} & 0 \\
0 & 0 & 0 & -\frac{\Xi}{2}\eta^{ade}\eta^{il} \\
-\frac{\Xi}{2}\eta^{abd}\eta^{il} & 0 & 0 & 0 \\
0 & \frac{\Xi}{2}\eta^{abd}\eta^{il} & 0 & 0
\end{array}
\right)\delta^{3}(x-y),
\end{eqnarray}
and the inverse of $C_{\alpha \beta}$ is given by 
\begin{eqnarray}
C^{-1}_{\alpha \beta} &=& \left(
\begin{array}{cccc}
0  & 0 & -\frac{2}{\Xi}\eta_{ade}\eta_{il} & 0 \\
0 & 0 & 0 & \frac{2}{\Xi}\eta_{ade}\eta_{il} \\
\frac{1}{\Xi}\eta_{abd}\eta_{il} & 0 & 0 & 0 \\
0 & -\frac{1}{\Xi}\eta_{abd}\eta_{il} & 0 & 0
\end{array}
\right)\delta^{3}(x-y).
\end{eqnarray}
Hence, the Dirac brackets between two functionals, namely $F(q, p)$ and $G(q,p)$,  is defined by 
\begin{eqnarray}\label{db}
\left\lbrace F,G  \right\rbrace _{D}\equiv \left\lbrace F,G \right\rbrace -\int dudv\left\lbrace F, \Gamma^{\alpha}(u) \right\rbrace C_{\alpha\beta}^{-1}(u,v)\left\lbrace \Gamma^{\beta}(v), G \right\rbrace, 
\end{eqnarray}
where $\Gamma^{\alpha}$ represent  the second class constraints. Therefore, the Dirac brackets between the fields are given by 
\begin{eqnarray}
\left\lbrace A_{a0i}(x), p^{d0j}(y) \right\rbrace _{D} &=& \left\lbrace A_{a0i}(x), p^{d0j}(y) \right\rbrace = \frac{1}{2}\delta_{a}^{d}\delta_{i}^{j}\delta^{3}(x-y),\\
\left\lbrace \Upsilon_{ai}(x), \pi^{dj}(y) \right\rbrace _{D} &=& \left\lbrace \Upsilon_{ai}(x), \pi^{dj}(y) \right\rbrace = \frac{1}{2}\delta_{a}^{d}\delta_{i}^{j}\delta^{3}(x-y),\\
\left\lbrace T_{i}(x), \hat{T}^{j}(y) \right\rbrace _{D} &=& \left\lbrace T_{i}(x), \hat{T}^{j}(y) \right\rbrace = \frac{1}{2}\delta_{i}^{j}\delta^{3}(x-y),\\
\left\lbrace \Lambda_{i}(x), \hat{\Lambda}^{j}(y) \right\rbrace _{D} &=& \left\lbrace \Lambda_{a0i}(x), \hat{\Lambda}^{d0j}(y) \right\rbrace = \frac{1}{2}\delta_{i}^{j}\delta^{3}(x-y),\\
\left\lbrace \varsigma_{ai}(x), \hat{\varsigma}^{dj}(y) \right\rbrace _{D} &=& \left\lbrace \varsigma_{ai}(x), \hat{\varsigma}^{dj}(y) \right\rbrace = \delta_{a}^{d}\delta_{i}^{j}\delta^{3}(x-y),\\
\left\lbrace \chi_{ai}(x), \hat{\chi}^{dj}(y) \right\rbrace _{D} &=& \left\lbrace \chi_{ai}(x), \hat{\chi}^{dj}(y) \right\rbrace = \delta_{a}^{d}\delta_{i}^{j}\delta^{3}(x-y),\\
\left\lbrace B_{ab0i}(x), p^{de0j}(y) \right\rbrace _{D} &=& 0,\\
\left\lbrace B_{abi}(x), p^{dej}(y) \right\rbrace _{D} &=& 0,\\
\left\lbrace A_{a0i}(x), B_{de0j}(y) \right\rbrace _{D} &=& -\frac{1}{2\Xi}\eta_{ade}\eta_{ij}\delta^{3}(x-y),\\
\left\lbrace \Upsilon_{ai}(x), B_{dej}(y) \right\rbrace _{D} &=& \frac{1}{2\Xi}\eta_{ade}\eta_{ij}\delta^{3}(x-y),
\end{eqnarray}
In addition,   we can also calculate  the  Dirac brackets  by gauge fixing,  in this case we will use the temporal   gauge, namely, we take   
\begin{eqnarray}
T_{i}&\approx & 0,\\
\Lambda_{i}&\approx & 0,\\
\varsigma_{ai} &\approx &0,\\
\chi_{ai} &\approx & 0.
\end{eqnarray}
These conditions are  considered as a new set of  second class constraints, hence, now  there are the following 84 second class constraints 
\begin{eqnarray}
\Gamma_{1}^{a0i} &\equiv & p^{a0i} - \Xi\eta^{abc}B_{bc}^{\;\;\;0i} \approx 0,\nonumber \\
\Gamma_{2}^{ai} &\equiv & \pi^{ai} - \Xi\eta^{abc}B_{bc}^{\;\;\;i} \approx 0,\nonumber \\
\Gamma_{3}^{ab0i} &\equiv & p^{ab0i} \approx 0,\nonumber \\
\Gamma_{4}^{abi} &\equiv & p^{abi} \approx 0,\nonumber \\
\Gamma_{5i} &\equiv & T_{i} \approx 0,\nonumber \\
\Gamma_{6i} &\equiv & \Lambda_{i} \approx 0,\nonumber \\
\Gamma_{7ai} &\equiv & \varsigma_{ai} \approx 0,\nonumber \\
\Gamma_{8ai} &\equiv & \chi_{ai} \approx 0,\nonumber \\
\Gamma_{9i} &\equiv & \hat{T}_{i} \approx 0,\nonumber \\
\Gamma_{10i} &\equiv & \hat{\Lambda}_{i} \approx 0,\nonumber \\
\Gamma_{11ai} &\equiv & \hat{\varsigma}_{ai} \approx 0,\nonumber \\
\Gamma_{12ai} &\equiv & \hat{\chi}_{ai} \approx 0.
\label{144a}
\end{eqnarray}
In this manner, the matrix $C_{\alpha\beta} = \left\lbrace \Gamma^{\alpha}, \Gamma^{\beta} \right\rbrace $ whose entries are given by the Poisson brackets between the second class constraints (\ref{144a}) is given by 
\small{\begin{eqnarray}
C_{\alpha \beta} 
&=& \left(
\begin{array}{cccccccccccc}
0  & 0 & \frac{\Xi}{2}\eta^{ade}\eta^{il} & 0 & 0 & 0 & 0 & 0 & 0 & 0 & 0 & 0 \\
0 & 0 & 0 & -\frac{\Xi}{2}\eta^{ade}\eta^{il} & 0 & 0 & 0 & 0 & 0 & 0 & 0 & 0  \\
-\frac{\Xi}{2}\eta^{abd}\eta^{il} & 0 & 0 & 0 & 0 & 0 & 0 & 0 & 0 & 0 & 0 & 0 \\
0 & \frac{\Xi}{2}\eta^{abd}\eta^{il} & 0 & 0 & 0 & 0 & 0 & 0 & 0 & 0 & 0 & 0 \\
0 & 0 & 0 & 0 & 0 & 0 & 0 & 0 & \frac{1}{2}\delta_{i}^{l} & 0 & 0 & 0\\
0 & 0 & 0 & 0 & 0 & 0 & 0 & 0 & 0 & \frac{1}{2}\delta_{i}^{l} & 0 & 0\\
0 & 0 & 0 & 0 & 0 & 0 & 0 & 0 & 0 & 0 & \delta_{a}^{d}\delta_{i}^{l} & 0\\
0 & 0 & 0 & 0 & 0 & 0 & 0 & 0 & 0 & 0 & 0 & \delta_{a}^{d}\delta_{i}^{l}\\
0 & 0 & 0 & 0 & -\frac{1}{2}\delta_{l}^{i} & 0 & 0 & 0 & 0 & 0 & 0 & 0\\
0 & 0 & 0 & 0 & 0 & -\frac{1}{2}\delta_{l}^{i} & 0 & 0 & 0 & 0 & 0 & 0\\
0 & 0 & 0 & 0 & 0 & 0 & -\delta_{d}^{a}\delta_{l}^{i} & 0 & 0 & 0 & 0 & 0\\
0 & 0 & 0 & 0 & 0 & 0 & 0 & -\delta_{d}^{a}\delta_{l}^{i} & 0 & 0 & 0 & 0
\end{array}
\right)\delta^{3}(x-y), \nonumber  \\
\end{eqnarray}}
and its inverse 
\small{\begin{eqnarray}
C^{-1}_{\alpha \beta}&=& \left(
\begin{array}{cccccccccccc}
0  & 0 & -\frac{2}{\Xi}\eta_{ade}\eta_{il} & 0 & 0 & 0 & 0 & 0 & 0 & 0 & 0 & 0 \\
0 & 0 & 0 & \frac{2}{\Xi}\eta_{ade}\eta_{il} & 0 & 0 & 0 & 0 & 0 & 0 & 0 & 0  \\
\frac{1}{\Xi}\eta_{abd}\eta_{il} & 0 & 0 & 0 & 0 & 0 & 0 & 0 & 0 & 0 & 0 & 0 \\
0 & -\frac{1}{\Xi}\eta_{abd}\eta_{il} & 0 & 0 & 0 & 0 & 0 & 0 & 0 & 0 & 0 & 0 \\
0 & 0 & 0 & 0 & 0 & 0 & 0 & 0 & -2\delta_{l}^{i} & 0 & 0 & 0\\
0 & 0 & 0 & 0 & 0 & 0 & 0 & 0 & 0 & -2\delta_{l}^{i} & 0 & 0\\
0 & 0 & 0 & 0 & 0 & 0 & 0 & 0 & 0 & 0 & -\delta_{a}^{d}\delta_{l}^{i} & 0\\
0 & 0 & 0 & 0 & 0 & 0 & 0 & 0 & 0 & 0 & 0 & -\delta_{a}^{d}\delta_{l}^{i}\\
0 & 0 & 0 & 0 & 2\delta_{i}^{l} & 0 & 0 & 0 & 0 & 0 & 0 & 0\\
0 & 0 & 0 & 0 & 0 & 2\delta_{i}^{l} & 0 & 0 & 0 & 0 & 0 & 0\\
0 & 0 & 0 & 0 & 0 & 0 & \delta_{a}^{d}\delta_{i}^{l} & 0 & 0 & 0 & 0 & 0\\
0 & 0 & 0 & 0 & 0 & 0 & 0 & \delta_{a}^{d}\delta_{i}^{l} & 0 & 0 & 0 & 0
\end{array}
\right) \delta^{3}(x-y). \nonumber \\
\label{eq130a}
\end{eqnarray}}
In this manner, by using (\ref{eq130a}) we  construct the Dirac brackets given by   
\begin{eqnarray}
\left\lbrace A_{a0i}(x), p^{d0j}(y) \right\rbrace_{D} &=& \left\lbrace A_{a0i}(x), p^{d0j}(y) \right\rbrace = \frac{1}{2}\delta_{a}^{d}\delta_{i}^{j}\delta^{3}(x-y),\nonumber \\
\left\lbrace \Upsilon_{ai}(x), \pi^{dj}(y) \right\rbrace_{D} &=& \left\lbrace \Upsilon_{ai}(x), \pi^{dj}(y) \right\rbrace = \frac{1}{2}\delta_{a}^{d}\delta_{i}^{j}\delta^{3}(x-y), \nonumber\\
\left\lbrace T_{i}(x), \hat{T}^{j}(y) \right\rbrace_{D} &=& 0, \nonumber\\
\left\lbrace \Lambda_{i}(x), \hat{\Lambda}^{j}(y) \right\rbrace_{D} &=& 0, \nonumber\\
\left\lbrace \varsigma_{ai}(x), \hat{\varsigma}^{dj}(y) \right\rbrace_{D} &=& 0, \nonumber\\
\left\lbrace \chi_{ai}(x), \hat{\chi}^{dj}(y) \right\rbrace_{D} &=& 0, \nonumber\\
\left\lbrace B_{ab0i}(x), p^{de0j}(y) \right\rbrace_{D} &=& 0, \nonumber\\
\left\lbrace B_{abi}(x), p^{dej}(y) \right\rbrace_{D} &=& 0, \nonumber\\
\left\lbrace A_{a0i}(x), B_{de0j}(y) \right\rbrace_{D} &=& -\frac{1}{2\Xi}\eta_{ade}\eta_{ij}\delta^{3}(x-y), \nonumber\\
\left\lbrace \Upsilon_{ai}(x), B_{dej}(y) \right\rbrace_{D} &=& \frac{1}{2\Xi}\eta_{ade}\eta_{ij}\delta^{3}(x-y),
\end{eqnarray}
where we can observe that the Dirac brackets and the generalized FJ brackets coincide to each other. 
\section{ Conclusions and prospects}
In this paper, the symplectic analysis of a four-dimensional $BF$ theory has been performed. We  reported the complete set of  FJ constraints and we  observe that there are present less constraints than in Dirac's method.  Furthermore,  we have carried  out the counting of physical degrees of freedom concluding that the theory is a topological one as expected. In addition,  we have used a temporal gauge in order to obtain a symplectic tensor, then the quantization  brackets of FJ were obtained. On the other hand, a pure canonical analysis has been performed. Under a laborious work, we have reported the complete set of first class and second class constraints, the algebra between the constraints is  in full agreement with the canonical rules; then  using a temporal gauge  the Dirac brackets  were computed. The FJ and Dirac's brackets coincide to each other, thus we can conclude that the FJ is more economical  than  Dirac framework.  Of course, if in the Dirac approach are introduced the Dirac brackets and   the second class constraints  are considered as strong equations, then the FJ and Dirac's constraints coincide to each other. Finally we would like  to comment  that in this work we provide   the necessary tools for studying  in alternative way the $BF$ formulations of gravity such as that reported in \cite{20}. In fact, in that work the canonical formulation of $BF$ gravity has been  performed, and will be interesting to study that theory by using   the ideas of the symplectic formalism of FJ. All these ideas are in progress and will be the subject of forthcoming works. 
\newline
\newline
\newline
\noindent \textbf{Acknowledgements}\\[1ex]
This work was supported by CONACyT under Grant No.CB-$2014$-$01/ 240781$. We would like to R. Cartas-Fuentevilla for discussion on the subject and reading of the manuscript.\\

\end{document}